\documentclass[11pt,onecolumn,letter]{IEEEtran}

\usepackage{amsmath}
\usepackage{amssymb}
\usepackage{amsfonts}
\usepackage{setspace}
\usepackage{cite}
\usepackage{ifthen}
\usepackage{epsfig}
\usepackage{array}
\usepackage{caption}
\usepackage{subfig}
\usepackage{color}
\usepackage[usenames,dvipsnames,table]{xcolor}
\usepackage{xspace}
\usepackage{multicol}
\usepackage{mathdots}
\usepackage{tabu}
\usepackage{colortbl}
\usepackage{mathtools}

\usepackage{placeins}

\definecolor{light-gray}{gray}{0.95}

\newtheorem{thm}{Theorem}

\newtheorem{cor}[thm]{Corollary}

\newtheorem{example}{Example}
\newtheorem{lemma}{Lemma}

\newcommand{\beq}{\begin{equation}}
\newcommand{\eeq}{\end{equation}}
\newcommand{\bea}{\begin{eqnarray}}
\newcommand{\eea}{\end{eqnarray}}
\newcommand{\beg}{\begin{example}\it\xspace}
\newcommand{\eeg}{\end{example}\xspace}

\newcommand{\msg}{\mathbf{u}}
\newcommand{\msga}{\mathbf{a}}
\newcommand{\msgb}{\mathbf{b}}
\newcommand{\msgc}{\mathbf{c}}
\newcommand{\msgd}{\mathbf{d}}
\newcommand{\msge}{\mathbf{e}}
\newcommand{\msgf}{\mathbf{f}}
\newcommand{\msgg}{\mathbf{g}}
\newcommand{\msgh}{\mathbf{h}}
\newcommand{\msgy}{\mathbf{y}}
\newcommand{\msgz}{\mathbf{z}}

\newcommand{\msgsymba}{a}
\newcommand{\msgsymbb}{b}
\newcommand{\msgsymbc}{c}
\newcommand{\msgsymbd}{d}
\newcommand{\msgsymbe}{e}
\newcommand{\msgsymbf}{f}
\newcommand{\msgsymbg}{g}
\newcommand{\msgsymbh}{h}
\newcommand{\p}{\mathbf{p}}
\newcommand{\psymb}{p}
\newcommand{\q}{\mathbf{q}}
\newcommand{\pa}{x}
\newcommand{\pb}{y}
\newcommand{\s}{\mathbf{v}}
\newcommand{\stripetype}{substripes\xspace}
\newcommand{\singstripetype}{substripe\xspace}
\newcommand{\stripesize}{number of substripes\xspace}
\newcommand{\normalnode}{\textnormal{Node}\xspace}
\newcommand{\access}{read\xspace}
\newcommand{\accessing}{reading\xspace}
\newcommand{\accessed}{{\access}\xspace}

\renewcommand{\arraystretch}{1.4}
\title{A Piggybacking Design Framework for Read-and Download-efficient Distributed Storage Codes}

\author{K.~V.~Rashmi, Nihar~B.~Shah, Kannan~Ramchandran, {\em Fellow, IEEE}\\Department of Electrical Engineering and Computer Sciences\\University of California, Berkeley.\\\{rashmikv,\,nihar,\,kannanr\}@eecs.berkeley.edu}

\begin{document}
\thispagestyle{empty}
\maketitle


\begin{figure*}[b!]
\vspace{-.1cm}
\medmuskip = .5\medmuskip
\thinmuskip = .3\thinmuskip
\thickmuskip = .5\thickmuskip
\centering
\subfloat{\hspace{-.4cm}
\begin{tabular}{c}
\multicolumn{1}{c}{}\\
\normalnode 1  \\
\normalnode 2  \\
\normalnode 3  \\
\normalnode 4  \\
\normalnode 5  \\
\normalnode 6\\
\end{tabular}
}\hspace{-.3cm}\setcounter{subfigure}{0}
\subfloat[]{
\begin{tabular}{|c|c|}
 \hline
\multicolumn{2}{|c|}{An MDS Code}\\
\hline 
$a_1$ & $b_1$ \\
\hline 
$a_2$ & $b_2$ \\
\hline 
$a_3$ & $b_3$ \\
\hline 
$a_4$ & $b_4$ \\
\hline 
$\sum_{i=1}^{4} a_i$ & $\sum_{i=1}^{4} b_i$ \\
\hline 
$\sum_{i=1}^{4} i a_i $ & $\sum_{i=1}^{4} i b_i$\\
\hline 
\end{tabular}\label{fig:RS}
}
\subfloat[]{
\begin{tabular}{|c|c|}
 \hline
\multicolumn{2}{|c|}{Intermediate Step}\\
\hline 
$a_1$ & $b_1$ \\
\hline 
$a_2$ & $b_2$ \\
\hline 
$a_3$ & $b_3$ \\
\hline 
 $a_4$ & $b_4$ \\
\hline 
$\sum_{i=1}^{4} a_i$ & $\sum_{i=1}^{4} b_i$ \\
\hline 
$\sum_{i=1}^{4} i a_i $ & \cellcolor{black!7} $\sum_{i=1}^{4} i b_i +\sum_{i=1}^{2} i a_i $ \\
\hline 
\end{tabular}\label{fig:RS_pb}
}
\subfloat[]{
\begin{tabular}{|c|c|}
 \hline
\multicolumn{2}{|c|}{Piggybacked Code}\\
\hline 
$a_1$ & $b_1$ \\
\hline 
$a_2$ & $b_2$ \\
\hline 
$a_3$ & $b_3$ \\
\hline 
$a_4$ & $b_4$ \\
\hline 
$\sum_{i=1}^{4} a_i$ & $\sum_{i=1}^{4} b_i$ \\
\hline 
 \cellcolor{black!7} $\sum_{i=3}^{4} i a_i - \sum_{i=1}^{4} i b_i $ & $\sum_{i=1}^{4} i b_i +\sum_{i=1}^{2} i a_i $ \\
\hline 
\end{tabular}\label{fig:RS_modpb}
}
\caption{An example illustrating efficient repair of systematic nodes using the piggybacking framework. Two instances of a $(6,4)$ MDS code are piggybacked to obtain a new $(6,4)$ MDS code that achieves $25\%$ savings in data-\access and download in the repair of any systematic node. A highlighted cell indicates a modified symbol.}
\label{tab:pbtoy}\raggedright
\end{figure*} 

\thispagestyle{empty}
\begin{abstract}
We present a new \textit{piggybacking} framework for designing distributed storage codes that are efficient in data-\access and download required during node-repair. We illustrate the power of this framework by constructing classes of explicit codes that entail the smallest data-\access and download for repair among all existing solutions for three important settings:
(a) codes meeting the constraints of being Maximum-Distance-Separable (MDS), high-rate and having a small \stripesize, arising out of practical considerations for implementation in data centers, (b) binary MDS codes for all parameters where binary MDS codes exist, (c) MDS codes with the smallest repair-locality.  In addition, we employ this framework to enable efficient repair of parity nodes in existing codes that were originally constructed to address the repair of only the systematic nodes. The basic idea behind our framework is to take multiple instances of existing codes and add carefully designed functions of the data of one instance to the other. Typical savings in data-\access during repair is $25\%$ to $50\%$ depending on the choice of the code parameters.
\end{abstract}

\begin{figure*}[t!]
\medmuskip = .5\medmuskip
\thinmuskip = .3\thinmuskip
\thickmuskip = .5\thickmuskip
\centering
\subfloat{
\begin{tabular}{c}
\normalnode 1  \\
\normalnode 2  \\
\normalnode 3  \\
\normalnode 4  \\
\normalnode 5  \\
\normalnode 6\\
\end{tabular}
}\hspace{-.3cm}
\subfloat{
\begin{tabular}{|c|c|c|c|}
\hline 
$a_1$ & $b_1$ & $c_1$ & $d_1$\\
\hline 
$a_2$ & $b_2$  & $c_2$ & $d_2$\\
\hline 
$a_3$ & $b_3$  & $c_3$ & $d_3$\\
\hline 
$a_4$ & $b_4$  & $c_4$ & $d_4$\\
\hline 
$\sum_{i=1}^{4} a_i$ & $\sum_{i=1}^{4} b_i$ & \cellcolor{black!7} $\sum_{i=1}^{4} c_i$ + $\sum_{i=1}^{4} i b_i  +\sum_{i=1}^{2} i a_i $ & $\sum_{i=1}^{4} d_i$\\
\hline 
$\sum_{i=3}^{4} i a_i - \sum_{i=1}^{4} i b_i $ & $\sum_{i=1}^{4} i b_i +\sum_{i=1}^{2} i a_i $ & $\sum_{i=3}^{4} i c_i - \sum_{i=1}^{4} i d_i $ & $\sum_{i=1}^{4} i d_i +\sum_{i=1}^{2} i c_i $\\
\hline 
\end{tabular}
}
\caption{An example illustrating of the mechanism of repair of parity nodes under the piggybacking framework, using two instances of the code of Fig.~\ref{fig:RS_modpb}.
A shaded cell indicates a modified symbol.}
\label{fig:RS_paritypb}
\end{figure*} 

\section{Introduction}\label{sec:intro}
Distributed storage systems today are increasingly employing erasure codes for data storage, since erasure codes provide much better storage efficiency and reliability as compared to replication-based schemes~\cite{hdfs_codes_blog,ford2010availability,cleversafe_erasure_codes}. Frequent failures of individual storage nodes in these systems mandate schemes for efficient repair of failed nodes. In particular, upon failure of a node, it is replaced by a new node, which must obtain the data that was previously stored in the failed node by \accessing and downloading data from the remaining nodes. Two primary metrics that determine the efficiency of repair are the amount of data \accessed at the remaining nodes (termed \textit{data-\access}\!) and the amount of data downloaded from them (termed \textit{data-download} or simply the \textit{download}).


In this paper, we present a new framework, which we call the \textit{piggybacking} framework, for design of repair-efficient storage codes. In a nutshell, this framework considers multiple instances of an existing code, and the piggybacking operation adds (carefully designed) functions of the data of one instance to the other. We design these functions with the goal of reducing the data-\access and download requirements during repair. Piggybacking preserves many of the properties of the underlying code such as the minimum distance and the field of operation. 


We need to introduce some notation and terminology at this point. Let $n$ denote the number of (storage) nodes and assume that the nodes have equal storage capacities. The data to be stored across these nodes is termed the \textit{message}. A \textit{Maximum-Distance-Separable (MDS)} code is associated to another parameter $k$: an $[n,\ k]$ MDS code guarantees that the message can be recovered from \textit{any} $k$ of the $n$ nodes, and requires a storage capacity of $\frac{1}{k}$ of the size of the message at every node. It follows that an MDS code can tolerate the failure of \textit{any} $(n-k)$ of the nodes without suffering any permanent data-loss. A \textit{systematic} code is one in which $k$ of the nodes store parts of the message without any coding. These $k$ nodes are termed the systematic nodes and the remaining $(n-k)$ nodes are termed the parity nodes. We denote the number of parity nodes by $r=(n-k)$. We shall assume without loss of generality that in a systematic code, the first $k$ nodes are systematic. 
The \textit{\stripesize }of a (vector) code is defined as the length of the vector of symbols that a node stores in a single instance of the code.

The piggybacking framework offers a rich design space for constructing codes for various different settings. We illustrate the power of this framework by providing the following four classes of explicit code constructions in this paper.

\paragraph*{(Class $1$) A class of codes meeting the constraints of being MDS, high-rate, and having a small \stripesize, with the smallest known average data-\access for repair} 
A major component of the cost of current day data-centers which store enormous amounts of data is the storage hardware. This makes it critical for any storage code to minimize the storage space utilization. In light of this, it is important for the erasure code employed to be MDS and have a high-rate (i.e., a small storage overhead). In addition, practical implementations also mandate a small \stripesize. There has recently been considerable work on the design of distributed storage codes with efficient data-\access during repair~\cite{ourAllerton,ourProductMatrix,tamo2011mds, cadambe2011permutation,khan2012rethinking,wang2011codes,jiekak2012system,rawat2012optimal,shum2012functional,rouayheb2010fractional,olmez2012repairable,han2013update,gaston2012quasi,sasidharan2013high,YunDimKanJournal,ourITW, oggier2011self,gopalan2011locality,papailiopoulos2012locally,kamath2012codes,ourAllertonJournal,arrayrepair2_dimakis, xiang2010optimal}. However, to the best of our knowledge, the only explicit codes that meet the aforementioned requirements are the Rotated-RS~\cite{khan2012rethinking} codes and the (repair-optimized) EVENODD~\cite{EVENODD,arrayrepair2_dimakis} and RDP~\cite{arraycodes5,xiang2010optimal} codes.  
Moreover, Rotated-RS codes exist only for $r \in \{2,\ 3\}$ and $k \leq 36$; the (repair-optimized) EVENODD and RDP codes exist only for $r=2$. Through our piggybacking framework, we construct a class of codes that are MDS, high-rate, have a small \stripesize, and require the least  amount of data-\access and download for repair among all other known codes in this class. An appealing feature of our codes is that they support all values of the system parameters $n$ and $k$. 

\paragraph*{(Class $2$) Binary MDS codes with the lowest known average data-\access for repair, for all parameters where binary MDS codes exist} 
Binary MDS codes are extensively used in disk arrays~\cite{EVENODD,arraycodes5}.  Through our piggybacking framework, we construct binary MDS codes that require the lowest known average data-\access for repair among all existing binary MDS codes~\cite{EVENODD,blaum1996mds,arraycodes5,xiang2010optimal,arrayrepair2_dimakis,khan2012rethinking}. Furthermore, unlike the other codes and repair algorithms~\cite{EVENODD,blaum1996mds,arraycodes5,xiang2010optimal,arrayrepair2_dimakis,khan2012rethinking} in this class, the codes constructed here also optimize the repair of parity nodes (along with that of systematic nodes). Our codes support all the parameters for which binary MDS codes are known to exist.


\paragraph*{(Class $3$) Efficient repair MDS codes with smallest possible repair-locality} Repair-locality is the number of nodes that need to be \accessed during repair of a node. 
While several recent works~\cite{oggier2011self,gopalan2011locality,papailiopoulos2012locally,kamath2012codes} present codes optimizing on locality, these codes are not MDS and hence require additional storage overhead for the same reliability levels as MDS codes. In this paper, we present MDS codes with efficient repair properties that have the smallest possible repair-locality for an MDS code.

\paragraph*{(Class $4$) A method of reducing data-\access and download for repair of parity nodes in existing codes that address only the repair of systematic nodes} The problem of efficient node-repair in distributed storage systems has attracted considerable attention in the recent past. However, many of the codes proposed~\cite{khan2012rethinking,cadambe2011permutation,ourITW, wang2011codes,papailiopoulos2011repair} have algorithms for efficient repair of \textit{only} the systematic nodes, and require the download of the entire message for repair of any parity node. In this paper, we employ our piggybacking framework to enable efficient repair of parity nodes in these codes, while also retaining the efficiency of repair of systematic nodes. The corresponding piggybacked codes enable an average saving of $25\%$ to $50\%$ in the amount of download and \access required for repair of parity nodes.

The following examples highlight the key ideas behind the piggybacking framework.
\beg\label{ex:toy_sys}
This example illustrates one method of piggybacking for reducing data-\access during systematic node repair. Consider two instances of a $(6,4)$ MDS code as shown in Fig.~\ref{fig:RS}, with the $8$ message symbols $\{a_i\}_{i=1}^{4}$ and $\{b_i\}_{i=1}^{4}$ (each column of Fig.~\ref{fig:RS} depicts a single instance of the code). One can verify that the message can be recovered from the data of any $4$ nodes. The first step of piggybacking involves adding $\sum_{i=1}^{2}i a_i$ to the second symbol of node $6$ as shown in Fig.~\ref{fig:RS_pb}. The second step in this construction involves subtracting the second symbol of node $6$ in the code of Fig.~\ref{fig:RS_pb} from its first symbol. The resulting code is shown in Fig.~\ref{fig:RS_modpb}. This code has $2$ \stripetype (the number of columns in Fig.~\ref{fig:RS_modpb}).

We now present the repair algorithm for the piggybacked code of Fig.~\ref{fig:RS_modpb}. Consider the repair of node $1$. Under our repair algorithm, the symbols $b_2, \ b_3, \ b_4$ and $\sum_{i=1}^4 b_i$ are download from the other nodes, and $b_1$ is decoded. In addition, the second symbol $(\sum_{i=1}^{4} i b_i +\sum_{i=1}^{2} i a_i )$ of node $6$ is downloaded. Subtracting out the components of $\{b_i\}_{i=1}^{4}$ gives the piggyback $\sum_{i=1}^{2} i a_i$. Finally, the symbol $a_2$ is downloaded from node $2$ and subtracted to obtain $a_1$. Thus, node $1$ is repaired by \accessing only $6$ symbols which is $75\%$ of the total size of the message. Node $2$ can be repaired in a similar manner. Repair of nodes $3$ and $4$ follows on similar lines except that the first symbol of node $6$ is \accessed instead of the second. 

The piggybacked code is MDS, and the entire message can be recovered from any $4$ nodes as follows. If node $6$ is one of these four nodes, then add its second symbol to its first, to recover the code of Fig.~\ref{fig:RS_pb}. Now, the decoding algorithm of the original code of Fig,~\ref{fig:RS} is employed to first recover $\{a_i\}_{i=1}^{4}$, which then allows for removal of the piggyback $(\sum_{i=1}^{2} i a_i)$ from the second substripe, making the remainder identical to the code of Fig.~\ref{fig:RS_modpb}.
\eeg

\beg\label{ex:toy_par}
This example illustrates the use of piggybacking to reduce data-\access during the repair of parity nodes. The code depicted in Fig.~\ref{fig:RS_paritypb} takes two instances of the code of Fig.~\ref{fig:RS_modpb}, and adds the second symbol of node $6$, $(\sum_{i=1}^{4} i b_i+\sum_{i=1}^{2} i a_i)$ (which belongs to the first instance), to the third symbol of node $5$ (which belongs to the second instance). This code has $4$ \stripetype (the number of columns in Fig.~\ref{fig:RS_paritypb}). In this code, repair of the second parity node involves downloading $\{a_i,c_i,d_i\}_{i=1}^{4}$ and the modified symbol $(\sum_{i=1}^{4} c_i+ \sum_{i=1}^{4} i b_i+\sum_{i=1}^{2} i a_i)$, using which the data of node $6$ can be recovered. The repair of the second parity node thus requires \access and download of only $13$ symbols instead of the entire message of size $16$. The first parity is repaired by downloading all $16$ message symbols. Observe that in the code of Fig.~\ref{fig:RS_modpb}, the first symbol of node $5$ is not used for repair of any of the systematic nodes. Thus the modification in Fig.~\ref{fig:RS_paritypb} does not change the algorithm or the efficiency of the repair of systematic nodes. The code retains its MDS property: the entire message can be recovered from any $4$ nodes by first decoding $\{a_i,b_i\}_{i=1}^{4}$ using the decoding algorithm of the code of Fig.~\ref{fig:RS}, which then allows for removal of the piggyback $(\sum_{i=1}^{4} i b_i+\sum_{i=1}^{2} i a_i)$ from the second instance, making the remainder identical to the code of Fig.~\ref{fig:RS}.
\eeg

Our piggybacking framework, enhances existing codes by adding piggybacks from one instance onto the other. The design of these piggybacks determine the properties of the resulting code. In this paper, we provide a few designs of piggybacking and specialize it to existing codes to obtain the four specific classes mentioned above. This framework, while being powerful, is also simple, and easily amenable for code constructions in other settings and scenarios.


The rest of the paper is organized as follows. Section~\ref{sec:framework} introduces the general piggybacking framework. Sections~\ref{sec:mthd1} and~\ref{sec:mthd2} then present code designs and repair-algorithms based on this framework, special cases of which result in classes $1$ and $2$ discussed above. Section~\ref{sec:mthd3} provides piggyback design which result in low-repair locality along with low data-\access and download. Section~\ref{sec:compare} provides a comparison of these codes and various other codes in the literature. Section~\ref{sec:regen} demonstrates the use of piggybacking to enable efficient parity repair in existing codes that were originally constructed for repair of only the systematic nodes. Section~\ref{sec:conclusions} draws conclusions. 

Readers interested only in repair locality may skip Sections~\ref{sec:mthd1} and~\ref{sec:mthd2}, and readers interested only in the mechanism of imbibing efficient parity-repair in existing codes optimized for systematic-repair may skip Sections~\ref{sec:mthd1}, \ref{sec:mthd2}, and~\ref{sec:mthd3} without any loss in continuity.


\section{The Piggybacking Framework}\label{sec:framework}
The piggybacking framework operates on an existing code, which we term the \textit{base code}. The choice of the base code is arbitrary. The base code is associated to $n$ encoding functions $\{f_i\}_{i=1}^{n}$: it takes the message $\msg$ as input and encodes it to $n$ coded symbols $\{f_1(\msg), \ldots , f_n(\msg)\}$. Node $i~(1\leq i \leq n)$ stores the data $f_i(\msg)$.

The piggybacking framework operates on multiple instances of the base code, and embeds information about one instance into other instances in a specific fashion. Consider $\alpha$ instances of the base code. The encoded symbols in $\alpha$ instances of the base code are\\

\begin{minipage}{\textwidth}
\begin{tabular}{c}
\normalnode 1  \\
\scriptsize$\vdots$\\
\normalnode n
\end{tabular}
\thinmuskip=.5\thinmuskip
\centering
  \begin{tabular}{|c|c|c|c|}
  \hline
$f_1(\msga)$ & $f_1(\msgb)$ & $\cdots$ & $f_1(\msgz)$ \\
\hline
 \scriptsize$\vdots$ & \scriptsize$\vdots$ & \scriptsize$\ddots$ & \scriptsize$\vdots$ \\
 \hline
 $f_n(\msga)$ & $f_n(\msgb)$ & $\cdots$ & $f_n(\msgz)$\\
\hline
\end{tabular}
\vspace{.145cm}
\end{minipage}
where $\msga, \ \ldots, \ \msgz$ are the (independent) messages encoded under these $\alpha$ instances.

We shall now describe the piggybacking of this code. For every $i,~2\leq i \leq \alpha$, one can add an arbitrary function of the message symbols of all previous instances $\{1, \ldots, (i-1)\}$ to the data stored under instance $i$. These functions are termed \textit{piggyback} functions, and the values so added are termed \textit{piggybacks}. Denoting the piggyback functions by $g_{i,j} \ (i\in\{2,\ldots,\alpha\}, \ j\in\{1,\ldots,n\})$, the piggybacked code is thus:

\begin{minipage}{\textwidth}
\begin{tabular}{c}
\normalnode 1  \\
\scriptsize$\vdots$\\
\normalnode n
\end{tabular}
\thinmuskip=.5\thinmuskip
\centering
\begin{tabular}{|c|c|c|c|c|}
\hline
 $f_1(\msga)$ & $f_1(\msgb)+ g_{2,1}(\msga)$ & $f_1(\msgc)+ g_{3,1}(\msga,\msgb)$ &$\cdots$ & $f_1(\msgz)+ g_{\alpha ,1}(\msga,\ldots,\msgy) $ \\
\hline
\scriptsize$\vdots$ & \scriptsize$\vdots$ & \scriptsize$\vdots$ &\scriptsize$\ddots$ & \scriptsize$\vdots$ \\
\hline
$f_n(\msga)$ & $f_n(\msgb)+ g_{2,n}(\msga)$ &$ f_1(\msgc)+ g_{3,n}(\msga,\msgb)$  & $\cdots$ & $f_n(\msgz)+ g_{\alpha ,n}(\msga,\ldots,\msgy)$\\
\hline
\end{tabular}
\end{minipage}

The decoding properties (such as the minimum-distance or the MDS nature) of the base code are retained upon piggybacking. In particular, the piggybacked code allows for decoding of the entire message from any set of nodes from which the base code allowed decoding. To see this, consider any set of nodes from which the message can be recovered in the base code. Observe that the first column of the piggybacked code is identical to a single instance of the base code. Thus $\msga$ can be recovered directly using the decoding procedure of the base code. The piggyback functions $\{g_{2,i}(\msga)\}_{i=1}^{n}$ can now be subtracted from the second column. The remainder of this column is precisely another instance of the base code, allowing recovery of $\msgb$. Continuing in the same fashion, for any instance $i~(2\leq i \leq n)$, the piggybacks (which are always a function of previously decoded instances $\{1,\ldots,i-1\}$) can be subtracted out to obtain the base code of that instance which can be decoded.

The decoding properties of the code are thus not hampered by the choice of the piggyback functions $g_{i,j}$'s. This allows for flexibility in the choice of the piggyback functions, and these need to be picked cleverly to achieve the desired goals (such as efficient repair, which is the focus of this paper).

The piggybacking procedure described above was followed in Example~\ref{ex:toy_sys} to obtain the code of Fig.~\ref{fig:RS_pb} from Fig.~\ref{fig:RS}. Subsequently, in Example~\ref{ex:toy_par}, this procedure was followed again to obtain the code of Fig.~\ref{fig:RS_paritypb} from Fig.~\ref{fig:RS_modpb}.

The piggybacking framework also allows any invertible linear transformation of the data stored in any individual node. In other words, each node of the piggybacked code (e.g., each row in Fig.~\ref{fig:RS_pb}) can separately undergo a invertible transformation. Clearly, any invertible transformation of data within the nodes does not alter the decoding capabilities of the code, i.e., the message can still be recovered from any set of nodes from which it could be recovered in the base code. In Example~\ref{ex:toy_sys}, the code of Fig.~\ref{fig:RS_modpb} is obtained from Fig.~\ref{fig:RS_pb} via an invertible transformation of the data of node $6$.


The following theorem formally proves that piggybacking does not reduce the amount of information stored in any subset of nodes. 
\begin{thm}\label{thm:framework}
Let $U_1,\ldots,U_{\alpha}$ be random variables corresponding to the messages associated to the $\alpha$ instances of the base code. For $i\in \{1,\ldots,n\}$, let  $X_i$ denote the data stored in node $i$ under the base code. Let $Y_i$ denote the encoded symbols stored in node $i$ under the piggybacked version of that code. Then for any subset of nodes $S \subseteq \{1,\ldots,n\}$,
\beq I\left(\left\lbrace Y_i \right \rbrace_{i \in S};U_1,\ldots,U_{\alpha}\right) \geq  I\left(\left\lbrace X_i \right \rbrace_{i \in S};U_1,\ldots,U_{\alpha}\right)~.  \eeq
\end{thm}
The proof of this theorem is provided in the appendix.

\begin{cor}
Piggybacking a code does not decrease its minimum distance; piggybacking an MDS code preserves the MDS property.
\end{cor}



\paragraph*{Notational Conventions} For simplicity of exposition, we shall assume throughout this section that the base codes are linear, scalar, MDS and systematic. Using vector codes (such as EVENODD or RDP) as base codes is a straightforward extension. The base code operates on a $k$-length message vector, with each symbol of this vector drawn from some finite field. The number of instances of the base code during piggybacking is denoted by $\alpha$, and $\{\msga,\msgb,\ldots\}$ shall denote the $k$-length message vectors corresponding to the $\alpha$ instances.  Since the code is systematic, the first $k$ nodes store the elements of the message vector. We use $\p_1,\ldots, \p_r$ to denote the $r$ encoding vectors corresponding to the $r$ parity symbols, i.e., if $\msga$ denotes the $k$-length message vector then the $r$ parity nodes under the base code store $\p_1^T \msga, \ldots,\p_r^T \msga$. 

The transpose of a vector or a matrix will be indicated by a superscript $^{T}$. Vectors are assumed to be column vectors. For any vector $\mathbf{v}$ of length $\kappa$, we denote its $\kappa$ elements as $\mathbf{v} = [{v}_{1} \ \cdots \ {v}_{\kappa}]^T$, and if the vector itself has an associated subscript then we its elements as $\mathbf{v}_i = [{v}_{i,1} \ \cdots \ {v}_{i,\kappa}]^T$.

Each of the explicit codes constructed in this paper possess the property that the repair of any node entails \accessing of only as much data as what has to be downloaded.~\footnote{In general, the amount of download lower bounds the amount of \access, and the download could be strictly smaller if a node passes a (non-injective) function of the data that it stores.} This property is called \textit{repair-by-transfer}~\cite{ourAllertonJournal}. Thus the amounts of data-\access and download are equal under our codes, and hence we shall use the same notation $\gamma$ to denote both these quantities. 


\FloatBarrier
\section{Piggybacking Design 1}\label{sec:mthd1}
In this section, we present our first design of piggyback functions and associated repair algorithms. 
This design allows one to reduce data-read and download during repair while having a small number of substripes. For instance, when the number of substripes is is small as $2$, we can achieve a $25$ to $35\% $ savings during repair of systematic nodes. We shall first present the piggyback design for optimizing the repair of systematic nodes, and then move on to the repair of parity nodes.

\subsection{Efficient repair of systematic nodes} \label{subsec:method1_sys}
This design operates on $\alpha=2$ instances of the base code. We first partition the $k$ systematic nodes into $r$ sets, $S_1,\ldots,S_r$. of equal size (or nearly equal size if $k$ is not a multiple of $r$). For ease of understanding, let us assume that $k$ is a multiple of $r$, which fixes the size of each of these sets as $\frac{k}{r}$. Then, let $S_1=\{1,\ldots,\frac{k}{r}\}$, $S_2=\{\frac{k}{r}+1,\ldots,\frac{2k}{r}\}$ and so on, with $S_i = \{\frac{(i-1)k}{r}+1,\ldots,\frac{ik}{r}\}$ for $i=1,\ldots,r$. 

Define the following $k-$length vectors:

\newcommand{\zeros}{0 ~~\cdots~~ 0}
\newcommand{\zerosleft}{0~\cdots~~}
\newcommand{\zerosright}{~~ \cdots ~0}
\begin{tabular}{>{$}l<{$}>{$}c<{$}>{$}c<{$}>{$}c<{$}>{$}c<{$}>{$}c<{$}>{$}c<{$}>{$}c<{$}>{$}l<{$}}
\thinmuskip=0\thinmuskip
\q_{2}&=&[&\psymb_{r,1}\,\cdots\, \psymb_{r,\frac{k}{r}} & \zerosleft &\cdots&\cdots& \zerosright&]^T\\
\q_{3}&=&[& \zeros &\psymb_{r,\frac{k}{r}+1}\, \cdots\, \psymb_{r,\frac{2k}{r}} & \zerosleft &\cdots&\zerosright&]^T\\
 &\vdots&&&&&&\\
\q_{r}&=&[&\zerosleft&\cdots&\zerosright&\psymb_{r,\frac{k}{r}(r-2)+1} ~ \cdots ~\psymb_{r,\frac{k}{r}(r-1)} & \zeros&]^T~\\
\q_{r+1}&=&[&\zerosleft&\cdots&\cdots&\zerosright&\psymb_{r,\frac{k}{r}(r-1)+1} ~ \cdots ~\psymb_{r,k}&]^T~.\\
\multicolumn{1}{l}~ \hspace{-.5cm}Also, let&&&\\
\s_{r} &=&\multicolumn{4}{l}{$\p_{r}-\q_{r}$}\\
&=&[&\psymb_{r,1}~~\cdots&\cdots &\cdots~~\psymb_{r,\frac{k}{r}(r-2)}&\zeros &\psymb_{r,\frac{k}{r}(r-1)+1}\, \cdots \,\psymb_{r,k} &]^T~.
\end{tabular}

\noindent Note that each element $\psymb_{i,j}$ is non-zero since the base code is MDS. We shall use this property during repair operations.

The base code is piggybacked in the following manner:\\

\begin{minipage}{\textwidth}
\begin{tabular}{c}
\normalnode 1  \\
\scriptsize$\vdots$\\
\normalnode k\\
\normalnode k+1  \\
\normalnode k+2 \\
\scriptsize$\vdots$\\
\normalnode k+r
\end{tabular}
\centering
\begin{tabular}{|c|c|c|}
\hline
$\msgsymba_{1}$  &$\msgsymbb_{1}$ \\
\hline
\scriptsize$\vdots$ &\scriptsize$\vdots$ \\
\hline
$\msgsymba_{k}$ &  $\msgsymbb_{k}$ \\
\hline
$\mathbf{p}_1^T \msga$ & $\mathbf{p}_1^T \msgb$ \\
\hline
$\mathbf{p}_2^T \msga$ &  $\mathbf{p}_2^T \msgb + \mathbf{q}_{2}^T \msga$ \\
\hline
\scriptsize$\vdots$ & \scriptsize$\vdots$ \\
\hline
$\mathbf{p}_r^T \msga$ & $\mathbf{p}_r^T \msgb + \mathbf{q}_{r}^T \msga$\\
\hline
\end{tabular}
\vspace{.3cm}
\end{minipage}

\noindent Fig.~\ref{fig:RS_pb} depicts an example of such a piggybacking.

We shall now perform an invertible transformation of the data stored in node $(k+r)$. In particular, the first symbol of node $(k+r)$ in the code above is replaced with the difference of this symbol from its second symbol, i.e., node $(k+r)$ now stores

\begin{minipage}{\textwidth}
\centering
\begin{tabular}{c}
\normalnode k+r
\end{tabular}
\begin{tabular}{|c|c|}
\hline
 $\s_{r}^T \msga - \p_r^T \msgb $ & $\mathbf{p}_r^T \msgb + \mathbf{q}_{r}^T \msga$\\
\hline
\end{tabular}
\vspace{.3cm}
\end{minipage}
The other symbols in the code remain intact. This completes the description of the encoding process.

Next, we present the algorithm for repair of any systematic node $\ell~(\in\{1,\ldots,k\})$. This entails recovery of the two symbols $\msgsymba_{\ell}$ and $\msgsymbb_{\ell}$ from the remaining nodes.

\textit{Case 1 ($\ell \notin S_r$):} Without loss of generality let $\ell \in S_1$. The $k$ symbols $\{\msgsymbb_{1},\ldots,\msgsymbb_{\ell-1},\msgsymbb_{\ell+1},\ldots,\msgsymbb_{k},\ \mathbf{p}_1^T \msgb\}$ are downloaded from the remaining nodes, and the entire vector $\msgb$ is decoded (using the MDS property of the base code). 
It now remains to recover $\msgsymba_{\ell}$. Observe that the $\ell^\textrm{th}$ element of $\q_{2}$ is non-zero. The symbol $(\p_2^T \msgb + \q_{2}^T \msga)$ is downloaded from node $(k+2)$, and since $\msgb$ is completely known, $\p_2^T\msgb$ is subtracted from the downloaded symbol to obtain the piggyback $\q_2^T\msga$. The symbols $\{\msgsymba_{i}\}_{i\in S_1\backslash\{\ell\}}$ are also downloaded from the other systematic nodes in set $S_1$. The specific (sparse) structure of $\q_{2}$ allows for recovering $\msgsymba_{\ell}$ from these downloaded symbols. Thus the total data-\access and download during the repair of node $\ell$ is $(k + \frac{k}{r})$ (in comparison, the size of the message is $2k$).  

\textit{Case $2$ ($S = S_{r}$):} As in the previous case, $\msgb$ is completely decoded  by downloading $\{\msgsymbb_{1},\ldots,\msgsymbb_{\ell-1},\msgsymbb_{\ell+1},\ldots,\msgsymbb_{k},\ \mathbf{p}_1^T \msgb\}$. The first symbol  $(\s_{r}^T \msga - \p_r^T \msgb)$ of node $(k+r)$ is downloaded. The second symbols $\{\mathbf{p}_i^T \msgb + \mathbf{q}_{i}^T \msga \}_i$ stored in the parities $i\in\{(k+2), \ldots, (k+r-1)\}$ are also downloaded, and are then subtracted from the first symbol of node $(k+r)$. This gives $(\q_{r+1}^T\msga + \mathbf{w}^T\msgb)$ for some vector $\mathbf{w}$. Using the previously decoded value of $\msgb$, $\mathbf{v}^T\msgb$ is removed to obtain $\q_{r+1}^T\msga$. Observe that the $\ell^\textrm{th}$ element of $\q_{r+1}$ is non-zero. The desired symbol $\msgsymba_{\ell}$ can thus be recovered by downloading $\{\msgsymba_{\frac{k}{r}(r-1)+1},\ldots,a_{\ell-1},a_{\ell+1},\ldots,\msgsymba_{k}\}$ from the other systematic nodes in $S_r$. The total data-\access and download required in recovering node $\ell$ is $(k + \frac{k}{r}+r-2)$.

Observe that the repair of systematic nodes in the last set $S_r$ requires more \access and download as compared to repair of systematic nodes in the other sets. Given this observation, we do not choose the sizes of the sets to be equal (as described previously), and instead optimize the sizes to minimize the average \access and download required. For $i=1,\ldots,r$, denoting the size of the set $S_i$ by $t_i$, the optimal sizes of the sets turn out to be
\bea
t_1 ~=~ \cdots ~=~ t_{r-1} & = & \left\lceil{\frac{k}{r} + \frac{r-2}{2r}}\right\rceil~:=~t, \\
t_r & = & k - (r-1)t~.
\eea
The amount of data \accessed and downloaded for repair of any systematic node in the first $(r-1)$ sets is $(k+t)$, and the last set is $(k+t_r+r-2)$. Thus, the average data-\access and download $\gamma_{1}^{\text{sys}}$ for repair of systematic nodes, as a fraction of the total number $2k$ of message symbols, is
\beq 
\gamma_{1}^{\text{sys}} = \frac{1}{2k^2}\left[\left(k-t_r\right)\left(k+t\right)+t_r\left(k + t_r+r-2\right) \right]~.\eeq
This quantity is plotted in Fig.~\ref{subfig:sys_compare} for various values of the system parameters $n$ and $k$.

\subsection{Reducing data-\access during repair of parity nodes} \label{subsec:method1_parity}
We shall now piggyback the code constructed in Section~\ref{subsec:method1_sys} to introduce efficiency in the repair of parity nodes, while also retaining the efficiency in the repair of systematic nodes. Observe that in the code of Section~\ref{subsec:method1_sys}, the first symbol of node $(k+1)$ is never \accessed for repair of any systematic node. We shall add piggybacks to this unused parity symbol to aid in the repair of other parity nodes. 

This design employs $m$ instances of the piggybacked code of Section~\ref{subsec:method1_sys}. The \stripesize in the resultant code is thus $2m$. The choice of $m$ can be arbitrary, and higher values of $m$ result in greater repair-efficiency. For every instance $i\in \{2,4,\ldots,2m-2\}$, the $(r-1)$ parity symbols in nodes $(k+2)$ to $(k+r)$ are summed up. The result is added as a piggyback to the $(i+1)^\textrm{th}$ symbol of node $(k+1)$. The resulting code, when $m=2$, is shown below.

\begin{minipage}{\textwidth}
\centering
\begin{tabular}{c}
\normalnode 1  \\
\scriptsize$\vdots$\\
\normalnode k\\
\normalnode k+1  \\
\normalnode k+2 \\
\scriptsize$\vdots$\\
\normalnode k+r-1\\
\normalnode k+r
\end{tabular}\hspace{-.1cm}
\begin{tabular}{|c|c|c|c|}
\hline
$\msgsymba_{1}$  &$\msgsymbb_{1}$ & $\msgsymbc_{1}$  &$\msgsymbd_{1}$ \\
\hline
\scriptsize$\vdots$ &\scriptsize$\vdots$&\scriptsize$\vdots$ &\scriptsize$\vdots$ \\
\hline
$\msgsymba_{k}$ &  $\msgsymbb_{k}$& $\msgsymbc_{k}$ &  $\msgsymbd_{k}$ \\
\hline
$\mathbf{p}_1^T \msga$ & $\mathbf{p}_1^T \msgb$&$\mathbf{p}_1^T \msgc + \sum_{i=2}^{r} (\mathbf{p}_i^T \msgb + \mathbf{q}_{i}^T \msga)$ & $\mathbf{p}_1^T \msgd$ \\
\hline
$\mathbf{p}_2^T \msga$ &  $\mathbf{p}_2^T \msgb + \mathbf{q}_{2}^T \msga$ &$\mathbf{p}_2^T \msgc$ &  $\mathbf{p}_2^T \msgd + \mathbf{q}_{2}^T \msgc$ \\
\hline
\scriptsize$\vdots$ & \scriptsize$\vdots$ & \scriptsize$\vdots$ & \scriptsize$\vdots$ \\
\hline
$\p_{r-\!1}^T \msga$ & $\mathbf{p}_{r-\!1}^T \msgb + \mathbf{q}_{r-\!1}^T \msga$& $\p_{r-\!1}^T \msgc$ & $\mathbf{p}_{r-\!1}^T \msgd + \mathbf{q}_{r-\!1}^T \msgc$\\
\hline
$\s_{r}^T \msga - \p_r^T \msgb$ & $\mathbf{p}_r^T \msgb + \mathbf{q}_{r}^T \msga$& $\s_{r}^T \msgc - \p_r^T \msgd$ & $\mathbf{p}_r^T \msgd + \mathbf{q}_{r}^T \msgc$\\
\hline
\end{tabular}
\vspace{.3cm}
\end{minipage}

\noindent This completes the encoding procedure. The code of Fig.~\ref{fig:RS_paritypb} is an example of this design.

As shown in Section~\ref{sec:framework}, the piggybacked code retains the MDS property of the base code. In addition, the repair of systematic nodes is identical to the that in the code of Section~\ref{subsec:method1_sys}, since the symbol modified in this piggybacking was never \accessed for the repair of any systematic node in the code of Section~\ref{subsec:method1_sys}.



We now present an algorithm for efficient repair of parity nodes under this piggyback design. The first parity node is repaired by downloading all $2mk$ message symbols from the systematic nodes. Consider repair of some other parity node, say node $\ell \in \{k+2,\ldots,k+r\}$. All message symbols $\msga,\msgc,\ldots$ in the odd substripes are downloaded from the systematic nodes. All message symbols of the last substripe (e.g., message $\msgd$ in the $m=2$ code shown above) are also downloaded from the systematic nodes. Further, the $\{3^\textrm{rd},5^\textrm{th},\ldots,(2m-1)^\textrm{th}\}$ symbols of node $(k+1)$ (i.e., the symbols that we modified in the piggybacking operation above) are also downloaded, and the components corresponding to the already downloaded message symbols are subtracted out. By construction, what remains in the symbol from substripe $i$ ($\in\{3,5,\ldots,2m-1\}$) is the piggyback. This piggyback is a sum of the parity symbols of the substripe $(i-1)$ from the last $(r-1)$ nodes (including the failed node). The remaining $(r-2)$ parity symbols belonging to each of the substripes $\{2,4,\ldots,2m-2\}$ are downloaded and subtracted out, to recover the data of the failed node. The procedure described above is illustrated via the repair of node $6$ in Example~\ref{ex:toy_par}.

The average data-\access and download $\gamma_{1}^{\text{par}}$ for repair of parity nodes, as a fraction of the total message symbols, is 
\beq\gamma_{1}^{\text{par}} = \frac{1}{2kr}\left[2k + (r-1) \left(\left(1+\frac{1}{m}\right)k + \left(1-\frac{1}{m}\right)(r-1)\right) \right] \nonumber~.\eeq
This quantity is plotted in Fig.~\ref{subfig:parity_compare} for various values of the system parameters $n$ and $k$.




\FloatBarrier
\section{Piggybacking Design 2}\label{sec:mthd2}
The design presented in this section provides a higher efficiency of repair as compared to the previous design. On the downside, it requires a larger \stripesize: the minimum \stripesize required under the design of Section~\ref{subsec:method1_sys} is $2$ and under that of Section~\ref{subsec:method1_parity} is $4$, while that required in the design of this section is $(2r-3)$. The following example illustrates this piggybacking design.


\beg
Consider some $(n=13, \ k=10 )$ MDS code as the base code, and consider $\alpha = (2r - 3) = 3$ instances of this code.  
Divide the systematic nodes into two sets of sizes $5$ each as $S_1 = \{1,\ldots, 5 \}$ and $S_2 = \{6,\ldots,10\}$. Define $10$-length vectors $\q_2$, $\s_2$, $\q_3$ and $\s_3$ as
\bea 
\q_2 &=& [\psymb_{2,1} \ \cdots \ \psymb_{2,5} ~~~~0~~~~~ \cdots~~~~~0] \nonumber\\
\s_2 &=& [0 \ ~~~~\cdots~~~~ 0 \ ~~~\psymb_{2,6} \ \cdots \ \psymb_{2,10}] \nonumber\\
\q_3 &=& [0 \ ~~~~\cdots~~~~ 0 \ ~~~\psymb_{3,6} \ \cdots \ \psymb_{3,10}]\nonumber\\
\s_3 &=& [\psymb_{3,1} \ \cdots \ \psymb_{3,5} ~~~~~0 ~~~~~ \cdots ~~~0] \nonumber
\eea

Now piggyback the base code in the following manner

\vspace{.2cm}
\begin{minipage}{\linewidth}
\centering
\begin{tabular}{c}
\normalnode \textnormal{1}  \\
\scriptsize$\vdots$\\
\normalnode \textnormal{10}\\
\normalnode \textnormal{11}  \\
\normalnode \textnormal{12} \\
\normalnode \textnormal{13}
\end{tabular}
\begin{tabular}{|c|c|c|}
\hline
$\msgsymba_{1}$ & $\msgsymbb_{1}$ & $\msgsymbc_{1} $\\
\hline
\scriptsize$\vdots$ & \scriptsize$\vdots$ & \scriptsize$\vdots$ \\
\hline
$\msgsymba_{10} $& $\msgsymbb_{10}$ & $\msgsymbc_{10}$ \\
\hline
$\p_1^T\msga$ & $\p_1^T\msgb$ & $\p_1^T\msgc$ \\
\hline
$\p_2^T\msga$ &  $\p_2^T\msgb +\q_2^T\msga$ & $\p_2^T\msgc+ \q_2^T\msgb+ \q_2^T\msga$ \\ 
\hline
$\p_3^T\msga$ & $\p_3^T\msgb +\q_3^T\msga$  & $\p_3^T\msgc+ \q_3^T\msgb+ \q_3^T\msga$ \\
\hline
\end{tabular}
\vspace{.2cm}
\end{minipage}

Next, we take invertible transformations of the (respective) data of nodes $12$ and $13$. The second symbol of node $i\in\{12,13\}$ in the new code is the difference between the second and the third symbols of node $i$ in the code above. The fact that $(\p_2-\q_2) = \s_2$ and $(\p_3-\q_3) = \s_3$ results in the following code

\vspace{.2cm}
\begin{minipage}{\linewidth}
\centering
\begin{tabular}{c}
\normalnode \textnormal{1}  \\
\scriptsize$\vdots$\\
\normalnode \textnormal{10}\\
\normalnode \textnormal{11}  \\
\normalnode \textnormal{12} \\
\normalnode \textnormal{13}
\end{tabular}
\begin{tabular}{|c|c|c|c|}
\hline
$\msgsymba_{1}$ & $\msgsymbb_{1}$ & $\msgsymbc_{1} $\\
\hline
\scriptsize$\vdots$ & \scriptsize$\vdots$ & \scriptsize$\vdots$ \\
\hline
$\msgsymba_{10} $& $\msgsymbb_{10}$ & $\msgsymbc_{10}$ \\
\hline
$\p_1^T\msga$ & $\p_1^T\msgb$ & $\p_1^T\msgc$ \\
\hline
$\p_2^T\msga$ &  $\s_2^T\msgb -\p_2^T\msgc$ & $\p_2^T\msgc+ \q_2^T\msgb+ \q_2^T\msga$ \\ 
\hline		
$\p_3^T\msga$ & $\s_3^T\msgb-\p_3^T\msgc$  & $\p_3^T\msgc+ \q_3^T\msgb+ \q_3^T\msga$ \\
\hline
\end{tabular} \vspace{.2cm}
\end{minipage}

\noindent 
This completes the encoding procedure.

We now present an algorithm for (efficient) repair of any systematic node, say node $1$. The  $10$ symbols $\{\msgsymbc_{2},\ldots,\msgsymbc_{10},\ \p_1^T\msgc\}$ are downloaded, and $\msgc$ is decoded. It now remains to recover $\msgsymba_{1}$ and $\msgsymbb_{1}$. The third symbol $(\p_2^T\msgc+ \q_2^T\msgb+ \q_2^T\msga)$ of node $12$ is downloaded and $\p_2^T\msgc$ subtracted out to obtain $(\q_2^T\msgb+ \q_2^T\msga)$. The second symbol $(\s_3^T\msgb-\p_3^T\msgc)$ from node $13$ is downloaded and $(-\p_3^T\msgc)$ is subtracted out from it to obtain $\s_3^T \msgb$. The specific (sparse) structure of $\q_2$ and $\s_3$ allows for decoding $\msgsymba_{1}$ and $\msgsymbb_{1}$ from $(\q_2^T\msgb+ \q_2^T\msga)$ and $\s_3^T \msgb$, by downloading and subtracting out $\{\msgsymba_{i}\}_{i=2}^5$ and $\{\msgsymbb_{i}\}_{i=2}^5$.  Thus, the repair of node $1$ involved \accessing and downloading $20$ symbols (in comparison, the size of the message is $k\alpha=30$). The repair of any other systematic node follows a similar algorithm, and results in the same amount of data-\access. 
\label{ex:mthd2}
\eeg

The general design is as follows. Consider $(2r-3)$ instances of the base code, and let $\msga_1,\ldots\msga_{2r-3}$ be the messages associated to the respective instances. First divide the $k$ systematic nodes into $(r-1)$ equal sets (or nearly equal sets if $k$ is not a multiple of $(r-1)$). Assume for simplicity of exposition that $k$ is a multiple of $(r-1)$. The first of the $\frac{k}{r-1}$ sets consist of the first $\frac{k}{r-1}$ nodes, the next set consists of the next $\frac{k}{r-1}$ nodes and so on. Define $k$-length vectors $\{\s_i,\ \hat{\s}_i\}_{i=2}^{r}$ as
\bea
\s_i &=& \msga_{r-1} + i \msga_{r-2} + i^2 \msga_{r-3} + \cdots + i^{r-2} \msga_1\nonumber\\
\hat{\s}_i &=& \s_i - \msga_{r-1}= i \msga_{r-2} + i^2 \msga_{r-3} + \cdots + i^{r-2} \msga_1~.\nonumber
\eea
Further, define $k$-length vectors $\{\q_{i,j}\}_{i=2,j=1}^{r,r-1}$ as
\[
\renewcommand{\arraystretch}{0}
\setlength{\tabcolsep}{0.1em}
\q_{i,j} = \left[ 
\thinmuskip=0\thinmuskip
\begin{tabular}{ccccccccc}
0&&&&&&&&\\
&$\scriptsize\ddots$&&&&&&&\\
&&0&&&&&&\\
&&&1&&&&&\\
&&&&$\scriptsize\ddots$&&&&\\
&&&&&1&&&\\
&&&&&&0&&\\
&&&&&&&$\scriptsize\ddots$&\\
&&&&&&&&0
\end{tabular}\right] \p_i
\]
where the positions of the ones on  the diagonal of the $(k \times k)$ diagonal matrix depicted correspond to the nodes in the $j^\textrm{th}$ group. It follows that
\[ \sum_{j=1}^{r-1} \q_{i,j} = \p_i \qquad \forall~i\in\{2,\ldots r\}~.\]

\noindent Parity node $(k+i)$, $i \in \{2,\ldots,r\}$,  is then piggybacked to store\\~\\
{
\thinmuskip=0\thinmuskip
\medmuskip=.3\medmuskip
\thickmuskip=.5\thickmuskip
\setlength{\tabcolsep}{0.1em}
\!\!\begin{tabular}{|>{$}c<{$}|>{$}c<{$}|>{$}c<{$}|>{$}c<{$}|>{$}c<{$}|>{$}c<{$}|>{$}c<{$}|>{$}c<{$}|>{$}c<{$}|>{$}c<{$}|}
\hline
\p_i^T\msga_1 & \cdots& \p_i^T \msga_{r-2} &  \p_i^T \msga_{r-1} + \sum_{j=1, j \neq i-1}^{r-1} \q_{i,j}^T\hat{\s}_i & \p_i^T \msga_{r} + \q_{i, 1}^T \s_i &\cdots & \p_i^T \msga_{r+i-3} + \q_{i, i-2}^T \s_i & \p_i^T \msga_{r} + \q_{i, i}^T \s_i  & \cdots & \p_i^T \msga_{2r-3} + \q_{i, r-1}^T \s_i \\
\hline
\end{tabular}
}

\noindent Following this, an invertible linear combination is performed at each of the nodes $\{k+2,\ldots,k+r\}$. The transform subtracts the last $(r-2)$ substripes from the $(r-1)^\textrm{th}$ substripe, following which the node $(k+i)$, $i \in \{2,\ldots,r\}$, stores\\~\\
{
\thinmuskip=0\thinmuskip
\medmuskip=.3\medmuskip
\thickmuskip=.5\thickmuskip
\setlength{\tabcolsep}{0.1em}
\!\!\begin{tabular}{|>{$}c<{$}|>{$}c<{$}|>{$}c<{$}|>{$}c<{$}|>{$}c<{$}|>{$}c<{$}|>{$}c<{$}|>{$}c<{$}|>{$}c<{$}|>{$}c<{$}|}
\hline
\p_i^T\msga_1 & \cdots& \p_i^T \msga_{r-2} &  \q_{i,i-1}^T \msga_{r-1} - \sum_{j=r}^{2r-3} \p_i^T \msga_{j}  & \p_i^T \msga_{r} + \q_{i, 1}^T \s_i &\cdots & \p_i^T \msga_{r+i-3} + \q_{i, i-2}^T \s_i & \p_i^T \msga_{r} + \q_{i, i}^T \s_i  & \cdots & \p_i^T \msga_{2r-3} + \q_{i, r-1}^T \s_i \\
\hline
\end{tabular}
}

Let us now see how repair of a systematic node is performed. Consider repair of node $\ell$. First, from nodes $\{1,\ldots,k+1\}\backslash\{\ell\}$, all the data in the last $(r-2)$ substripes is downloaded and the data $\msga_{r},\ldots,\msga_{2r-3}$ is recovered. This also provides us with the desired data $\{a_{r,\ell},\ldots,a_{2r-3,\ell}\}$. Next, observe that in each parity node $\{k+2,\ldots,k+r\}$, there is precisely one `$\q$' vector that has a non-zero first component. From each of these nodes, the symbol having this vector is downloaded, and the components along $\{\msga_{r},\ldots,\msga_{2r-3}\}$ are subtracted out. Further, we download  all symbols from all other systematic nodes in the same set as node $\ell$, and subtract this out from the previously downloaded symbols. This leaves us with $(r-1)$ independent linear combinations of $\{\msgsymba_{1,\ell},\ldots,\msgsymba_{r-1,\ell}\}$ from which the desired data is decoded.

When $k$ is not a multiple of $(r-1)$, the $k$ systematic nodes are divided into $(r-1)$ sets as follows. Let
\beq
t_{\ell} = \left\lfloor \frac{k}{r-1}\right\rfloor~, \quad \quad t_{h} = \left\lceil\frac{k}{r-1}\right\rceil~, \quad\quad t=(k - (r-1)t_{\ell})~.
\eeq
The first $t$ sets are chosen of size $t_h$ each and the remaining $(r-1-t)$ sets have size $t_{\ell}$ each. 
The systematic symbols in the first $(r-1)$ \stripetype are piggybacked onto the parity symbols (except the first parity) of the last $(r-1)$ stripes. For repair of any failed systematic node $\ell \in \{1,\ldots,k\}$, the last $(r-2)$ \stripetype are decoded completely by \accessing the remaining systematic and the first parity symbols from each. To obtain the remaining $(r-1)$ symbols of the failed node, the $(r-1)$ parity symbols that have piggyback vectors (i.e., $\q$'s and $\s$'s) with a non-zero value of the $\ell^\textrm{th}$ element are downloaded. By design, these piggyback vectors have non-zero components only along the systematic nodes in the same set as node $\ell$. Downloading and subtracting these other systematic symbols gives the desired data.  

The average data-\access  and download $\gamma_{2}^{\text{sys}}$ for repair of systematic nodes, as a fraction of the total message symbols $2k$, is 
\bea
 \gamma_{2}^{\text{sys}} & = & \frac{1}{2k^2} [ t ((r-2)k + (r-1)t_h) + (k-t)((r-2)k + (r-1)t_{\ell})]~.
\eea
This quantity is plotted in Fig.~\ref{subfig:sys_compare} for various values of the system parameters $n$ and $k$.

While we only discussed the repair of systematic nodes for this code, the repair of parity nodes can be made efficient by considering $m$ instances of this code. A procedure analogous to that described in Section~\ref{subsec:method1_parity} is followed, where the odd instances are piggybacked on to the succeeding even instances. As in Section~\ref{subsec:method1_parity}, higher a value of $m$ results in a lesser amount of data-\access and download for repair. In such a design, the average data-\access and download $\gamma_{2}^{\text{par}}$ for repair of parity nodes, as a fraction of the total message symbols, is 
\beq
\gamma_{2}^{\text{par}}  = \frac{1}{r} + \frac{r-1}{2r-3}\left[ (m+1)(r-2)k + (m-1)(r-2)(r-1) + \left\lceil\frac{m}{2}\right\rceil k + \left\lfloor \frac{m}{2} \right\rfloor (r-1)\right] ~.  
\eeq
This quantity is also plotted in Fig.~\ref{subfig:parity_compare} for various values of the system parameters $n$ and $k$.



\FloatBarrier
\section{Piggybacking Design 3}\label{sec:mthd3}
In this section, we present a piggybacking design to construct MDS codes with a primary focus on the \textit{locality} of repair. The locality of a repair operation is defined as the number of nodes that are contacted during the repair operation. The codes presented here perform the efficient repair of any systematic node with the smallest possible locality for any MDS code, which is equal to $(k+1)$.~\footnote{A locality of $k$ is also possible, but this necessarily mandates the download of the entire data, and hence we do not consider this option.} The amount of \access and download is the smallest among all known MDS codes with this locality, when $(n-k)>2$. 


This design involves two levels of piggybacking, and these are illustrated in the following two example constructions. The first example considers $\alpha=2m$ instances of the base code and shows the first level of piggybacking, for any arbitrary choice of $m >1$. Higher values of $m$ result in repair with a smaller \access and download. The second example uses two instances of this code and adds the second level of piggybacking. We note that this design deals with the repair of only the systematic nodes.


\beg \label{ex:method3}
\newcommand{\meg}{2}
\newcommand{\nex}{11}
\newcommand{\keg}{8}
\newcommand{\alphaeg}{4}
Consider any $(n=\nex, \ k=\keg )$ MDS code as the base code, and take $\alphaeg$ instances of this code. Divide the systematic nodes into two sets as follows, $S_1= \{1,2,3, 4\},\ S_2=\{5, 6,7, 8\}$. We then add the piggybacks as shown in Fig.~\ref{fig:method3}. Observe that in this design, the piggybacks added to an even \singstripetype is a function of symbols in its immediately previous (odd) \singstripetype from only the systematic nodes in the first set $S_1$, while the piggybacks added to an odd \singstripetype are functions of symbols in its immediately previous (even) \singstripetype from only the systematic nodes in the second set $S_2$. 

\begin{figure*}[tbp]
\begin{center}
\begin{tabu}{c}
\rowfont{\color{blue}}
\normalnode \textnormal{1}  \\
\scriptsize$\vdots$\\
\rowfont{\color{blue}}
\normalnode \textnormal{4}\\
\rowfont{\color{green}}
\normalnode \textnormal{5}  \\
\scriptsize$\vdots$\\
\rowfont{\color{green}}
\normalnode \textnormal{\keg}\\
\normalnode \textnormal{9}  \\
\normalnode \textnormal{10}  \\
\normalnode \textnormal{11} 
\end{tabu}
\begin{tabu}{|c|c|c|c|}
\hline
\rowfont{\color{blue}}
$\msgsymba_{1}$ & $\msgsymbb_{1}$ & $\msgsymbc_{1} $& $\msgsymbd_{1} $\\
\hline
\scriptsize$\vdots$ & \scriptsize$\vdots$ & \scriptsize$\vdots$ & \scriptsize$\vdots$  \\
\hline
\rowfont{\color{blue}}
 $\msgsymba_{4}$ & $\msgsymbb_{4}$ & $\msgsymbc_{4} $& $\msgsymbd_{4} $\\
\hline
\rowfont{\color{Green}}
$\msgsymba_{5}$ & $\msgsymbb_{5}$ & $\msgsymbc_{5} $& $\msgsymbd_{5} $\\
\hline
\scriptsize$\vdots$ & \scriptsize$\vdots$ & \scriptsize$\vdots$ & \scriptsize$\vdots$  \\
\hline
\rowfont{\color{Green}}
 $\msgsymba_{\keg} $& $\msgsymbb_{\keg}$ & $\msgsymbc_{\keg}$ & $\msgsymbd_{\keg}$  \\
\hline
$\p_1^T\msga$ & $\p_1^T\msgb$ & $\p_1^T\msgc$ & $\p_1^T\msgd$ \\
\hline	
 $\p_2^T\msga$ &  $\p_2^T\msgb \color{blue}{+\msgsymba_{1}+\msgsymba_{2}}$ & $\p_2^T\msgc \color{Green}{+\msgsymbb_{5}+\msgsymbb_{6}}$ & $\p_2^T\msgd \color{blue}{+\msgsymbc_{1}+\msgsymbc_{2}}$ \\ 
\hline
$\p_3^T\msga$ & $\p_3^T\msgb \color{blue}{+\msgsymba_{3}+\msgsymba_{4}}$  & $\p_3^T\msgc\color{Green}{+\msgsymbb_{7}+\msgsymbb_{8}}$ & $\p_3^T\msgd\color{blue}{+\msgsymbc_{3}+\msgsymbc_{4}}$ \\
\hline
\end{tabu} 
\caption{Example illustrating first level of piggybacking in design $3$. The piggybacks in the even \stripetype (in blue) are a function of only the systematic nodes $\{1,  \ldots, 4\} $ (also in blue), and the piggybacks in  odd \stripetype (in  green) are a function of only the systematic nodes $\{5,  \ldots, 8\} $ (also in green). This code requires an average data-\access and download of only $71\%$ of the message size for repair of systematic nodes.} 
\label{fig:method3}
\end{center}
\end{figure*}

We now present the algorithm for repair of any systematic node. First consider the repair of any systematic node $\ell\in \lbrace 1,\ldots,4\rbrace$ in the first set. For instance, say $\ell = 1$, then $\{\msgsymbb_{2},\ldots,\msgsymbb_{8},\ \p_1^T\msgb\}$ and $\{\msgsymbd_{2},\ldots,\msgsymbd_{8},\ \p_1^T\msgd\}$  are downloaded, and $\{ \msgb, \ \msgd\}$ (i.e., the messages in the even substripes) are decoded. It now remains to recover the symbols $\msgsymba_{1}$ and $\msgsymbc_{1}$ (belonging to the odd \stripetype). The second symbol $(\p_2^T\msgb+a_1 + a_2)$ from node $10$ is downloaded and $\p_2^T\msgb$ subtracted out to obtain the piggyback $(a_1+a_2)$. Now $\msgsymba_{1}$ can be recovered by downloading and subtracting out $\msgsymba_{2}$. The fourth symbol from node $10$,  $(\p_2^T\msgd+ c_1+c_2)$, is also downloaded and $\p_2^T\msgd$ subtracted out to obtain the piggyback $(c_1+c_2)$. Finally, $\msgsymbc_{1}$ is recovered by downloading and subtracting out $\msgsymbc_{2}$. Thus, node $1$ is repaired by by \accessing a total of $20$ symbols (in comparison, the total total message size is $32$). The repair of node $2$ can be carried out in an identical manner. The two other nodes in the first set, nodes $3$ and $4$, can be repaired in a similar manner by \accessing  the second and fourth symbols of node $11$ which have their piggybacks. 
Thus, repair of any node in the first group requires \accessing and downloading a total of $20$ symbols.

Now we consider the repair of any node $\ell \in \lbrace 5,\ldots,8\rbrace$ in the second set $S_2$. For instance, consider $\ell=5$. The symbols $\left\lbrace \msgsymba_{1},\ldots,\msgsymba_{8},\ \p_1^T\msga\right\rbrace \backslash \lbrace \msgsymba_{5}\rbrace$, $\left\lbrace \msgsymbc_{1},\ldots,\msgsymbc_{8},\ \p_1^T\msgc\right\rbrace \backslash \lbrace \msgsymbc_{5}\rbrace$ and $\left\lbrace \msgsymbd_{1},\ldots,\msgsymbd_{8},\ \p_1^T\msgd\right\rbrace \backslash \lbrace \msgsymbd_{5}\rbrace$ are downloaded in order to decode $\msgsymba_{5}, \msgsymbc_{5},$ and $\msgsymbd_{5}$. From node $10$, the symbol $(\p_2^T\msgc+ b_5 + b_6)$ is downloaded and $\p_2^T\msgc$ is subtracted out. Then, $\msgsymbb_{5}$ is recovered by downloading and subtracting out $\msgsymbb_{6}$. Thus, node $5$ is recovered by \accessing a total of $26$ symbols. Recovery of other nodes in $S_2$ follows on similar lines. 

The average amount of data \accessed and downloaded during the recovery of systematic nodes is $23$, which is $71\%$ of the message size. A higher value of $m$ (i.e., a higher number of \stripetype) would lead to a further reduction in the \access and download (the last substripe cannot be piggybacked and hence mandates a greater \access and download; this is a boundary case, and its contribution to the overall \access reduces with an increase in $m$). 
\eeg
\FloatBarrier

\newcommand{\nforex}{13}
\newcommand{\kforeg}{10}
\begin{figure*}[btp]
\newcolumntype{g}{>{\columncolor{light-gray}}c}
\begin{center}
\newcommand{\figheight}{1.33in}
\resizebox{!}{\figheight}{
\begin{tabu}{c}
\rowfont{\color{blue}}
\normalnode \textnormal{1}  \\
\scriptsize$\vdots$\\
\rowfont{\color{blue}}
\normalnode \textnormal{4}\\
\rowfont{\color{green}}
\normalnode \textnormal{5}  \\
\scriptsize$\vdots$\\
\rowfont{\color{green}}
\normalnode \textnormal{8}\\
\rowfont{\color{red}}
\normalnode \textnormal{9}  \\
\rowfont{\color{red}}
\normalnode \textnormal{10}  \\
\normalnode \textnormal{11} \\
\normalnode \textnormal{12} \\
\normalnode \textnormal{13} 
\end{tabu}
}\!\!\!
\resizebox{!}{\figheight}{
\thinmuskip=0\thinmuskip
\medmuskip=0.2\medmuskip
\thickmuskip=0\thickmuskip
\begin{tabu}{|c|c|c|c|g|g|g|g|}
\hline
$\msgsymba_{1}$ & $\msgsymbb_{1}$ & $\msgsymbc_{1} $& $\msgsymbd_{1} $ & $\msgsymbe_{1}$ & $\msgsymbf_{1}$ & $\msgsymbg_{1} $& $\msgsymbh_{1} $\\
\hline
\scriptsize$\vdots$ & \scriptsize$\vdots$ & \scriptsize$\vdots$ & \scriptsize$\vdots$ & \scriptsize$\vdots$ & \scriptsize$\vdots$ & \scriptsize$\vdots$ & \scriptsize$\vdots$ \\
\hline
$\msgsymba_{4}$ & $\msgsymbb_{4}$ & $\msgsymbc_{4} $& $\msgsymbd_{4} $ & $\msgsymbe_{4}$ & $\msgsymbf_{4}$ & $\msgsymbg_{4} $& $\msgsymbh_{4} $\\
\hline
 $\msgsymba_{5}$ & $\msgsymbb_{5}$ & $\msgsymbc_{5} $& $\msgsymbd_{5} $ & $\msgsymbe_{5}$ & $\msgsymbf_{5}$ & $\msgsymbg_{5} $& $\msgsymbh_{5} $\\
\hline
\scriptsize$\vdots$ & \scriptsize$\vdots$ & \scriptsize$\vdots$ & \scriptsize$\vdots$ & \scriptsize$\vdots$ & \scriptsize$\vdots$ & \scriptsize$\vdots$ & \scriptsize$\vdots$ \\
\hline
 $\msgsymba_{8} $& $\msgsymbb_{8}$ & $\msgsymbc_{8}$ & $\msgsymbd_{8}$ & $\msgsymbe_{8} $& $\msgsymbf_{8}$ & $\msgsymbg_{8}$ & $\msgsymbh_{8}$ \\
\hline
$\msgsymba_{9} $& $\msgsymbb_{9}$ & $\msgsymbc_{9}$ & $\msgsymbd_{9}$ & $\msgsymbe_{9} $& $\msgsymbf_{9}$ & $\msgsymbg_{9}$ & $\msgsymbh_{9}$ \\
\hline
$\msgsymba_{\kforeg} $& $\msgsymbb_{\kforeg}$ & $\msgsymbc_{\kforeg}$ & $\msgsymbd_{\kforeg}$ & $\msgsymbe_{\kforeg} $& $\msgsymbf_{\kforeg}$ & $\msgsymbg_{\kforeg}$ & $\msgsymbh_{\kforeg}$ \\
\hline
 $\p_1^T\msga$ & $\p_1^T\msgb$ & $\p_1^T\msgc$ & $\p_1^T\msgd$& $\p_1^T\msge \color{red}{+\msgsymba_9 + \msgsymba_{10}}$ & $\p_1^T\msgf \color{red}{+\msgsymbb_9 + \msgsymbb_{10}}$ & $\p_1^T\msgg \color{red}{+\msgsymbc_9 + \msgsymbc_{10}}$ & $\p_1^T\msgh \color{red}{+\msgsymbc_9 + \msgsymbc_{10}}$ \\
\hline	
$\p_2^T\msga$ &  $\p_2^T\msgb \color{blue}{+\msgsymba_1 + \msgsymba_{2}}$ & $\p_2^T\msgc \color{Green}{+\msgsymbb_{5} + \msgsymbb_{6}}$ & $\p_2^T\msgd \color{blue}{+\msgsymbc_{1} + \msgsymbc_{2}}$ & $\p_2^T\msge$ &  $\p_2^T\msgf \color{blue}{+\msgsymbe_{1} + \msgsymbe_{2}}$ & $\p_2^T\msgg \color{Green}{+\msgsymbf_{5} + \msgsymbf_{6}}$ & $\p_2^T\msgh \color{blue}{+\msgsymbg_{1} + \msgsymbg_{2}}$ \\ 
\hline
$\p_3^T\msga$ & $\p_3^T\msgb \color{blue}{+\msgsymba_{3} + \msgsymba_{4}}$  & $\p_3^T\msgc\color{Green}{+\msgsymbb_{7} + \msgsymbb_{8}}$ & $\p_3^T\msgd\color{blue}{+\msgsymbc_{3} + \msgsymbc_{4}}$ & $\p_3^T\msge $ & $\p_3^T\msgf \color{blue}{+\msgsymbe_{3} + \msgsymbe_{4}}$  & $\p_3^T\msgg\color{Green}{+\msgsymbf_{7} + \msgsymbf_{8}}$ & $\p_3^T\msgh\color{blue}{+\msgsymbg_{3} + \msgsymbg_{4}}$ \\
\hline
\end{tabu} 
}
\caption{An example illustrating piggyback design $3$, with $k=10, \ n=13, \  \alpha = 8$. The piggybacks in the first parity node (in red) are functions of the data of nodes $\{8, \ 9\}$ alone. In the remaining parity nodes, the piggybacks in the even \stripetype (in blue) are functions of the data of nodes $\{1,  \ldots, 4\}$ (also in blue), and the piggybacks in the odd substripes (in green) are functions of the data of nodes $\{5,  \ldots, 8\}$ (also in green),  and the piggybacks in red (also in red). The piggybacks in nodes $12$ and $13$ are identical to that in Example~\ref{ex:method3} (Fig~\ref{fig:method3}). The piggybacks in node $11$ piggyback the first set of $4$ \stripetype (white background) onto the second set of \stripesize (gray background)}. 
\label{fig:method3b}
\end{center}
\end{figure*}

\beg
In this example, we illustrate the second level of piggybacking which further reduces the amount of data-\access during repair of systematic nodes as compared to Example~\ref{ex:method3}. Consider $\alpha = 8$ instances of an $(n=\nforex, \ k=\kforeg )$ MDS code. Partition the systematic nodes into three sets $S_1 =\{ 1, \ldots ,\ 4 \}$, $S_2 = \{5,\ldots,8\}$, $S_3=\{ 9, \ 10 \}$ (for readers having access to Fig.~\ref{fig:method3b} in color, these nodes are coloured blue, green, and red respectively). We first add piggybacks of the data of the first $8$ nodes onto the parity nodes $12$ and $13$ exactly as done in Example~\ref{ex:method3} (see Fig.~\ref{fig:method3b}). We now add piggybacks for the symbols stored in systematic nodes in the third set, i.e., nodes $9$ and $10$. To this end, we parititon the $8$ substripes into two groups of size four each (indicated by white and gray shades respectively in Fig.~\ref{fig:method3b}). The symbols of nodes $9$ and $10$ in the first four substripes are piggybacked onto the last four substripes of the first parity node, as shown in Fig.~\ref{fig:method3b} (in red color).


We now present the algorithm for repair of systematic nodes under this piggyback code. 
The repair algorithm for the systematic nodes $\{1,\ldots,8\}$ in the first two sets closely follows the repair algorithm illustrated in Example~\ref{ex:method3}. Suppose $\ell \in S_1$, say $\ell=1$. By construction, the piggybacks corresponding to the nodes in $S_1$ are present in the parities of even \stripetype. 
From the even \stripetype, the remaining systematic symbols, $\{ \msgsymbb_{i},\msgsymbd_{i},\msgsymbf_{i},\msgsymbh_{i}\}_{i=\{2,\ldots,10\}}$, and the symbols in the first parity, $\{\p_1^T\msgb,\ \p_1^T\msgd, \ \p_1^T\msgf+\msgsymbb_{9}+\msgsymbb_{10}, \ \p_1^T\msgh+\msgsymbd_{9}+\msgsymbd_{10}\}$, are downloaded. Observe that, the first two parity symbols downloaded do not have any piggybacks. Thus, using the MDS property of the base code, $\msgb$ and $\msgd$ can be decoded. This also allows us to recover $\p_1^T\msgf, \ \p_1^T\msgh$ from the symbols already downloaded. Again, using the MDS property of the base code, one recovers $\msgf$ and $\msgh$. It now remains to recover $\{a_1,\ c_1,\ e_1,\ g_1\}$. To this end, we download the symbols in the even \stripetype of node $12$, $\{\p_2^T\msgb {+\msgsymba_1 + \msgsymba_{2}},\ \p_2^T\msgd {+\msgsymbc_1 + \msgsymbc_{2}},\ \p_2^T\msgf {+\msgsymbe_1 + \msgsymbe_{2}},\ \p_2^T\msgh {+\msgsymbg_1 + \msgsymbg_{2}}\}$, which have piggybacks with the desired symbols. By subtracting out previously downloaded data, we obtain the piggybacks  $\{\msgsymba_1 + \msgsymba_{2}, \msgsymbc_1 + \msgsymbc_{2}, \msgsymbe_1 + \msgsymbe_{2},\msgsymbg_1 + \msgsymbg_{2}\}$. Finally, by downloading and subtracting $\msgsymba_2,\msgsymbc_2,\msgsymbe_2,\msgsymbg_2$, we recover $\msgsymba_1,\msgsymbc_1,\msgsymbe_1,\msgsymbg_1$. Thus, node $1$ is recovered by \accessing $48$ symbols, which is $60\%$ of the total message size. Observe that the repair of node $1$ was accomplished by downloading data from only $(k+1)=11$ other nodes. Every node in the first set can be repaired in a similar manner. Repair of the systematic nodes in the second set is performed in a similar fashion by utilizing the corresponding piggybacks, however, the total number of symbols \accessed is $64$ (since the last \singstripetype cannot be piggybacked; such was the case in Example~\ref{ex:method3} as well).

We now present the repair algorithm for systematic nodes $\{9,\ 10\}$ in the third set $S_3$. Let us suppose $\ell=9$. Observe that the piggybacks corresponding to node $9$ fall in the second group (i.e., the last four) of \stripetype. From the last four \stripetype, the remaining systematic symbols $\{ \msgsymbe_{i},\msgsymbf_{i},\msgsymbg_{i},\msgsymbh_{i}\}_{i=\{1,\ldots,8,10\}}$, and the symbols in the second parity $\{\p_1^T\msge, \ \p_1^T\msgf+\msgsymbe_{1}+\msgsymbe_{2}, \ \p_1^T\msgg+\msgsymbf_{1}+\msgsymbf_{2},\ \p_1^T\msgh+\msgsymbg_{1}+\msgsymbg_{2}, \}$ are downloaded. Using the MDS property of the base code, one recovers $\msge$, $\msgf$, $\msgg$ and $\msgh$. It now remains to recover $a_9$, $b_9$, $c_9$ and $d_9$. To this end, we download $\{\p_1^T\msge+\msgsymba_{9}+\msgsymba_{10}, \ \p_1^T\msgf+\msgsymbb_{9}+\msgsymbb_{10}, \ \p_1^T\msgg+\msgsymbc_{9}+\msgsymbc_{10},\ \p_1^T\msgh+\msgsymbd_{9}+\msgsymbd_{10} \}$ from node $11$. Subtracting out the previously downloaded data, we obtain the piggybacks $\{\msgsymba_{9}+\msgsymba_{10}, \ \msgsymbb_{9}+\msgsymbb_{10}, \ \msgsymbc_{9}+\msgsymbc_{10},\ \msgsymbd_{9}+\msgsymbd_{10} \}$.
Finally, by downloading and subtracting out $\{\msgsymba_{10},\msgsymbb_{10},\msgsymbc_{10},\msgsymbd_{10}\}$, we recover the desired data $\{\msgsymba_9,\msgsymbb_9,\msgsymbc_9,\msgsymbd_9\}$. Thus, node $9$ is recovered by \accessing and downloading $48$ symbols. Observe that the repair process involved reading data from only $(k+1)=11$ other nodes. Node $10$ is repaired in a similar manner. 
\label{ex:method3b}
\eeg 

For general values of the parameters, $n, \ k$, and $\alpha = 4m$ for some integer $m >1$, we choose the size of the three sets $S_1$, $S_2$, and $S_3$, so as to make the number of systematic nodes involved in each piggyback equal or nearly equal. Denoting the sizes of $S_1$, $S_2$ and $S_3$, by $t_1$, $t_2$, and $t_3$ respectively, this gives
\beq t_1 = \left\lceil \frac{1}{2r-1}\right\rceil~, \quad t_2 = \left\lceil \frac{r-1}{2r-1}\right\rceil~, \quad t_3 = \left\lfloor \frac{r-1}{2r-1}\right\rfloor~.
\eeq
Then the average data-\access and download $\gamma_3^{\text{sys}}$ for repair of systematic nodes, as a fraction of the total message symbols $4mk$, is
\bea \gamma_{3}^{\text{sys}} & = & \frac{1}{4mk^2} \left[ t_1 \left(\frac{k}{2}+\frac{t_1}{2}\right)  + t_2 \left(\frac{k}{2}+\frac{t_2}{2(r-1)} \right) + t_3 \left(\left(\frac{1}{2}+\frac{1}{m}\right)k +\left(\frac{1}{2}-\frac{1}{m}\right)\frac{t_3}{(r-1)}\right) \right]~. \eea
This quantity is plotted in Fig.~\ref{subfig:sys_compare} for various values of the system parameters $n$ and $k$.

\FloatBarrier
\section{Comparison of different codes}\label{sec:compare}
We now compare the average data-\access and download  entailed during repair under the piggyback constructions with various other storage codes in the literature.  As discussed in Section~\ref{sec:intro}, practical considerations in data centers require the storage codes to be MDS, high-rate, and have a small \stripesize. The table below compares different explicit codes designed for efficient repair, with respect to whether they are MDS or not, the parameters they support and the \stripesize. Shaded cells indicate a violation of the aforementioned requirements. The parameter $m$ associated to the piggyback codes can be chosen to have any value $m\geq1$. The base code for each of the piggyback constructions is a Reed-Solomon code~\cite{reed1960polynomial}.

\vspace{.3cm}
\begin{minipage}{.99\linewidth}
\setlength{\tabcolsep}{1em}
\centering
\begin{tabular}{|c|c|c|c|}
\hline
Code& MDS & $k,r$ supported & Number of substripes \\
\hline
High-rate Regenerating\cite{wang2011codes,cadambe2011permutation} & Y & $r\in\{2,3\}$ & \cellcolor{black!7} {$r^{\frac{k}{r}}$} \\
Product-Matrix MSR~\cite{ourProductMatrix} & Y & \cellcolor{black!7} {$r \geq k-1$} & $r$ \\
Local Repair~\cite{oggier2011self,gopalan2011locality,papailiopoulos2012locally,kamath2012codes} & \cellcolor{black!7} {N} & all & 1 \\
Rotated RS~\cite{khan2012rethinking} & Y & $r\in\{2,3\},k\leq 36$ & $2$\\
EVENODD, RDP~\cite{EVENODD,arraycodes5,arrayrepair2_dimakis,xiang2010optimal}&Y&$r=2$&$k$\\
Piggyback 1 & Y & all & $2m$ \\
Piggyback 2 & Y & $r\geq 3$ & $(2r-3)m$ \\
Piggyback 3 & Y & all & $4m$ \\
\hline
\end{tabular}\label{tab:compare}
\end{minipage}
\vspace{.2cm}

The piggyback, rotated-RS, (repair-optimized) EVENODD and RDP codes satisfy the desired conditions. 
Fig.~\ref{fig:compare} shows a plot comparing the repair properties of these codes. The plot corresponds to the \stripesize being $8$ in Piggyback 1 and Rotated-RS, $4(2r-3)$ in Piggyback 2, and $16$ in the Piggyback 3 codes. We observe from the plot that piggyback codes require a lesser (average) data-\access and download as compared to Rotated-RS, (repair-optimized) EVENODD and RDP. 
 
\newcommand{\figsize}{0.3}
\begin{figure*}
\centering
\subfloat[Systematic]{
\includegraphics[trim=2in 2.8in 2in 2.8in, width=\figsize\textwidth]{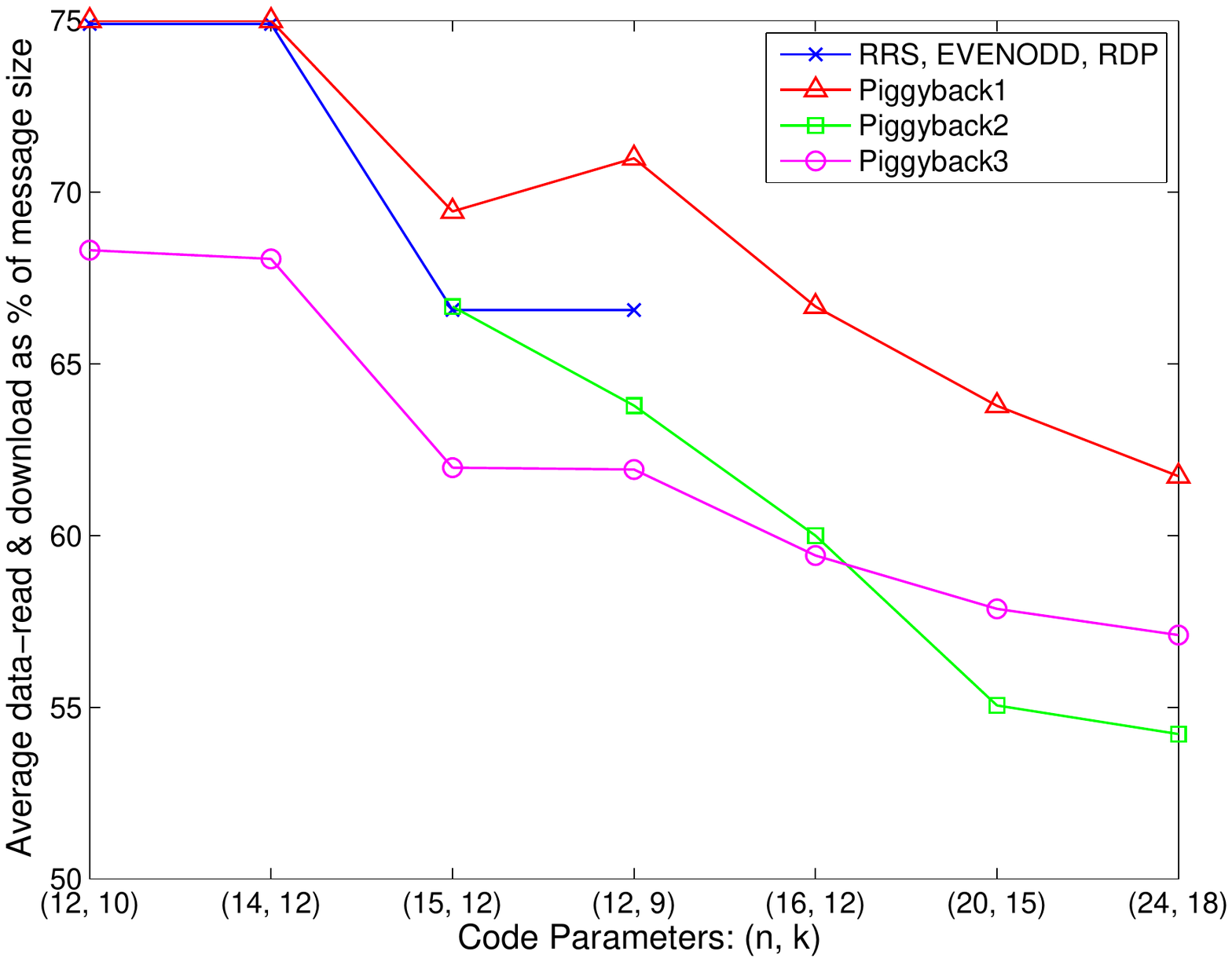}
\label{subfig:sys_compare}
}\qquad\qquad\qquad\qquad\qquad
\subfloat[Parity]{
\includegraphics[trim=2in 2.8in 2in 2.8in, width=\figsize\textwidth]{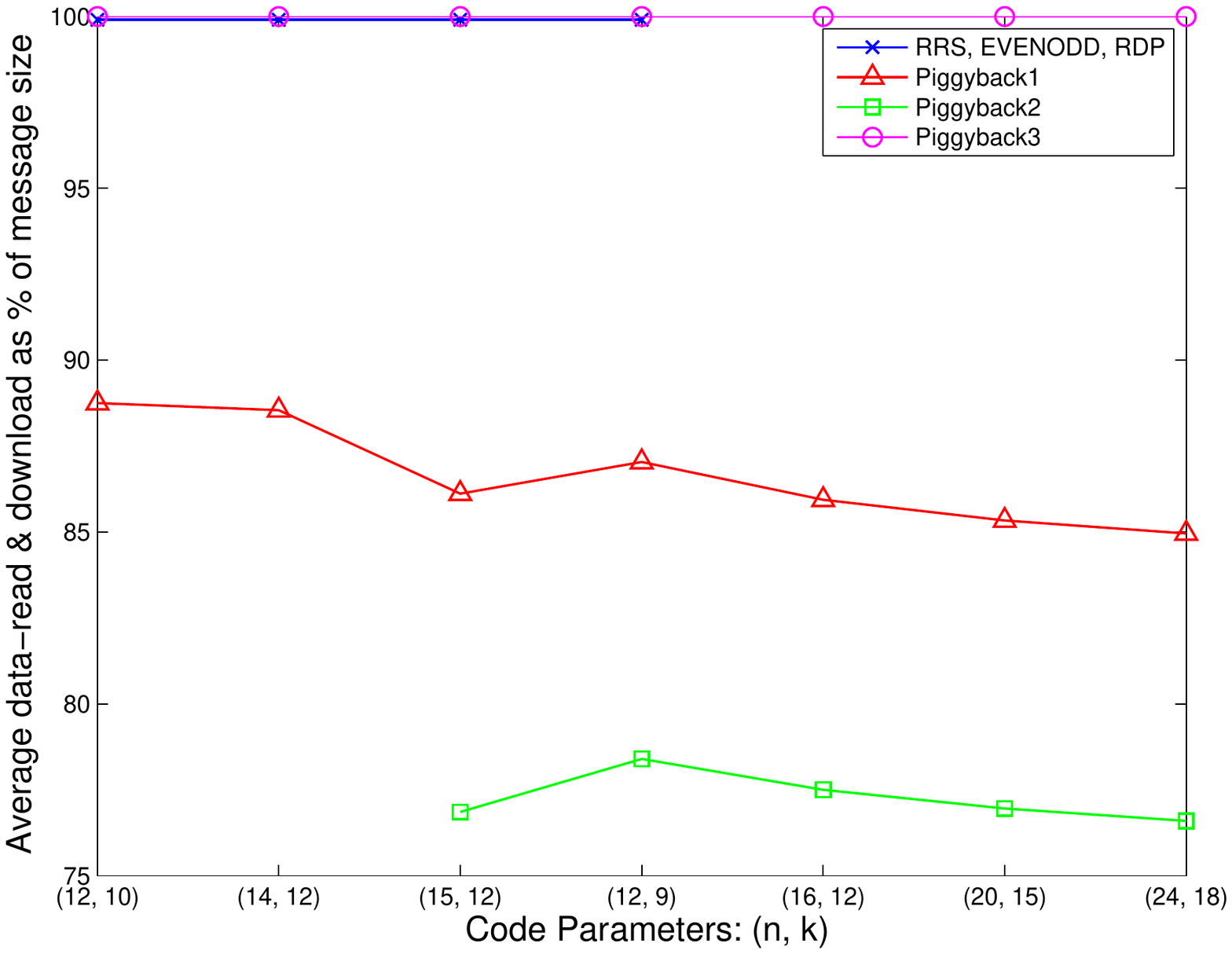}
\label{subfig:parity_compare}
}\\
\subfloat[Overall]{
\includegraphics[trim=2in 2.8in 2in 2.8in, width=\figsize\textwidth]{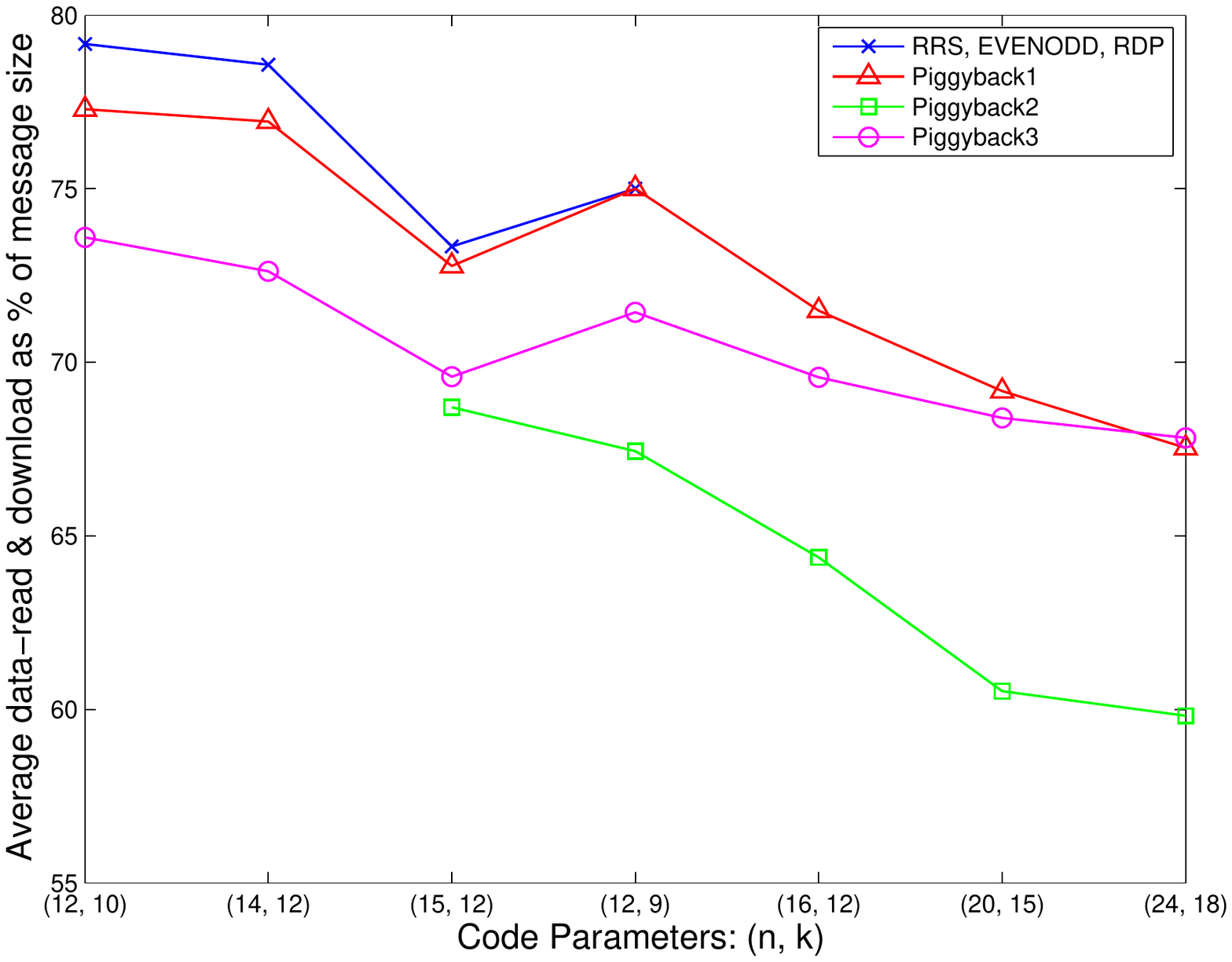}
}
\captionof{figure}{Average data-\access and download for repair of systematic, parity, and all nodes in the three piggybacking designs, Rotated-RS codes~\cite{khan2012rethinking}, and (repair-optimized) EVENODD and RDP codes~\cite{xiang2010optimal,arrayrepair2_dimakis}. The (repair-optimized) EVENODD and RDP codes exist only for $(n-k)=2$, and the data-\access and download required for repair are identical to that of a rotated-RS codes with the same parameters. While the savings plotted correspond to a relatively small number of substripes, an increase in this number improves the performance of the piggybacked codes.
}
\vspace{.7cm}
\label{fig:compare}
\end{figure*}

\section{Repairing parities in existing codes that address only systematic repair}\label{sec:regen}

Several codes proposed in the literature~\cite{khan2012rethinking,cadambe2011permutation,ourITW,papailiopoulos2011repair} can efficiently repair only the systematic nodes, and require the download of the entire message for repair of any parity node. In this section, we piggyback these codes to reduce the \access and download during repair of parity nodes, while also retaining the efficiency of repair of systematic nodes. This piggybacking design is first illustrated with the help of an example.

\beg\label{ex:regen}
\begin{figure*}
\setlength{\tabcolsep}{0.4em}
\medmuskip = .3\medmuskip
\thinmuskip = .3\thinmuskip
\centering
\subfloat{\hspace{-.5cm}
\begin{tabular}{c}
\normalnode 1  \\
\normalnode 2  \\
\shortstack{~\\~\\\normalnode 3\\~}  \\
\normalnode 4  \\
\end{tabular}
}\hspace{-.3cm}\setcounter{subfigure}{0}
\subfloat[An existing code~\cite{ourITW} originally designed to address repair of only systematic nodes]{
\begin{tabular}{|c|c|}
 \hline
$a_1$ & $b_1$ \\
\hline 
$a_2$ & $b_2$ \\
\hline 
\shortstack{~\\~\\$3a_1+2b_1+a_2$\\~} & \shortstack{~\\~\\$b_1+2a_2+3b_2$\\~} \\
\hline 
$3a_1+4b_1+2a_2$ & $b_1+2a_2+b_2$\\
\hline 
\end{tabular}\label{fig:regen_ex1}
}
\subfloat[Piggybacking to also optimize repair of parity nodes]{
\begin{tabular}{|c|c|c|c|}
\hline 
$a_1$ & $b_1$ & $c_1$ & $d_1$\\
\hline 
$a_2$ & $b_2$  & $c_2$ & $d_2$\\
\hline 
\shortstack{$3a_1+2b_1+a_2$\\~} & \shortstack{$b_1+2a_2+3b_2$\\~} & \shortstack{~\\$3c_1+2d_1+c_2$\\$\color{red}+(3a_1+4b_1+2a_2)$} & \shortstack{~\\$d_1+2c_2+3d_2$\\$\color{red}+(b_1+2a_2+b_2)$}\\
\hline 
$3a_1+4b_1+2a_2$ & $b_1+2a_2+b_2$ & $3c_1+4d_1+2c_2$ & $d_1+2c_2+d_2$\\
\hline 
\end{tabular}
\label{fig:regen_ex2}
}\hspace{-.3cm}\!\!
\label{fig:regen_ex}
\caption{An example illustrating piggybacking to perform efficient repair of the parities in an existing code that originally addressed the repair of only the systematic nodes. See Example~\ref{ex:regen} for more details.}
\end{figure*} 
Consider the code depicted in Fig.~\ref{fig:regen_ex1}, originally proposed in~\cite{ourITW}. This is an MDS code with parameters $(n=4,\ k=2)$, and the message comprises four symbols $a_1$, $a_2$, $b_1$ and $b_2$ over finite field $\mathbb{F}_5$. The code can repair any systematic node with an optimal data-\access and download.   Node $1$ is repaired by \accessing and downloading the symbols $a_2$, $(3a_1+2b_1+a_2)$ and $(3a_1+4b_1+2a_2)$ from nodes $2$, $3$ and $4$ respectively; node $2$ is repaired by \accessing and downloading the symbols $b_1$,  $(b_1+2a_2+3b_2)$ and $(b_1+2a_2+b_2)$ from nodes $1$, $3$ and $4$ respectively. The amount of data-\accessed and downloaded in these two cases are the minimum possible. However, under this code, the repair of parity nodes with reduced data-\access has not been addressed. 

In this example, we piggyback the code of Fig.~\ref{fig:regen_ex1} to enable efficient repair of the second parity node. In particular, we take two instances of this code and piggyback it in a manner shown in Fig.~\ref{fig:regen_ex2}. This code is obtained by piggybacking on the first parity symbol of the last two instance, as shown in Fig.~\ref{fig:regen_ex2}. In this piggybacked code, repair of systematic nodes follow the same algorithm as in the base code, i.e., repair of node $1$ is accomplished by downloading the first and third symbols of the remaining three nodes, while the repair of node $2$ is performed by downloading the second and fourth symbols of the remaining nodes. One can easily verify that the data obtained in each of these two cases is identical to what would have been obtained in the code of Fig.~\ref{fig:regen_ex1} in the absence of piggybacking. Thus the repair of the systematic nodes remains optimal. Now consider repair of the second parity node, i.e., node $4$. The code (Fig.~\ref{fig:regen_ex1}), as proposed in~\cite{ourITW}, would require \accessing $8$ symbols (which is the size of the entire message) for this repair. However, the piggybacked version of Fig.~\ref{fig:regen_ex2} can accomplish this task by \accessing and downloading only $6$ symbols: $c_1$, $c_2$, $d_1$, $d_2$, $(3c_1+2d_1+c_2+3a_1+4b_1+2a_2)$ and $(d_1+2c_2+3d_2+b_1+2a_2+b_2)$. Here, the first four symbols help in the recovery of the last two symbols of node $4$, $(3c_1+4d_1+2c_2)$ and $(d_1+2c_2+d_2)$. Further, from the last two downloaded symbols, $(3c_1+2d_1+c_2)$ and $(d_1+2c_2+3d_2)$ can be subtracted out (using the known values of $c_1$, $c_2$, $d_1$ and $d_2$) to obtain the remaining two symbols $(3a_1+4b_1+2a_2)$ and $(b_1+2a_2+b_2)$. Finally, one can easily verify that the MDS property of the code in Fig.~\ref{fig:regen_ex1} carries over to Fig.~\ref{fig:regen_ex2} as discussed in Section~\ref{sec:framework}.
\eeg

We now present a general description of this piggybacking design. We first set up some notation. Let us assume that the base code is a vector code, under which each node stores a vector of length $\mu$ (a scalar code is, of course, a special case with $\mu=1$). Let $\msga = [\msga_1^T~~\msga_2^T~\cdots~\msga_k^T]^T$ be the message, with systematic node $i~(\in \{1,\ldots,k\})$ storing the $\mu$ symbols $\msga_i^T$. Parity node $(k+j),~j \in \{1,\ldots,r\}$, stores the vector $\msga^T P_j$ of $\mu$ symbols for some $(k\mu \times \mu)$ matrix $P_j$. Fig.~\ref{fig:regen1} illustrates this notation using two instances of such a (vector) code.

We assume that in the base code, the repair of any failed node requires only linear operations at the other nodes. More concretely, for repair of a failed systematic node $i$, parity node $(k+j)$ passes $\msga^T P_j Q_{j}^{(i)}$ for some matrix $Q_{j}^{(i)}$. 

The following lemma serves as a building block for this design.
\begin{lemma}\label{lem:regen1}
Consider two instances of any base code, operating on messages $\msga$ and $\msgb$ respectively. Suppose there exist two parity nodes $(k+\pa)$ and $(k+\pb)$, a $(\mu \times \mu)$ matrix $R$, and another matrix $S$ such that 
\beq R Q_{\pa}^{(i)} =  Q_{\pb}^{(i)} S \qquad \forall~i\in \{1,\ldots,k\} \label{eq:regen}~.\eeq
Then, adding $\msga^T P_\pb R$ as a piggyback to the parity symbol $\msgb^T P_\pa$ of node $(k+\pa)$ (i.e., changing it from $\msgb^T P_\pa$ to $(\msgb^T P_\pa+\msga^T P_\pb R)$) does not alter the amount of \access or download required during repair of any systematic node. 
\end{lemma}
\begin{IEEEproof}
Consider repair of any systematic node $i \in\{1,\ldots,k\}$. In the piggybacked code, we let each node pass the same linear combinations of its data as it did under the base code. This keeps the amount of \access and download identical to the base code. Thus, parity node $(k+\pa)$ passes $\msga^T P_\pa Q_\pa^{(i)}$ and $(\msgb^T P_\pa+\msga^T P_\pb R)Q_\pa^{(i)}$, while parity node $(k+\pb)$ passes $\msga^T P_\pb Q^{(i)}_\pb$ and $\msgb^T P_\pb Q^{(i)}_\pb$. From~\eqref{eq:regen} we see that the data obtained from parity node $(k+\pb)$ gives access to $\msga^T P_\pb Q^{(i)}_\pb S = \msga^T P_\pb R Q_{\pa}^{(i)} $. This is now subtracted from the data downloaded from node $(k+\pa)$ to obtain $\msgb^T P_\pa Q_\pa^{(i)}$. At this point, the data obtained is identical to what would have been obtained under the repair algorithm of the base code, which allows the repair to be completed successfully. 
\end{IEEEproof}
An example of such a piggybacking is depicted in Fig.~\ref{fig:regen2}.

\begin{figure*}[tb!]
\centering
\subfloat[Two instances of the vector base code.]{
\begin{tabular}{c}
\normalnode 1  \\
$\vdots$\\
\normalnode k \\
\normalnode k+1\\
\normalnode k+2\\
$\vdots$\\
\normalnode k+r \\
\end{tabular}
\begin{tabular}{|c|c|}
\hline 
$\msga_1^T$ & $\msgb_1^T$ \\
\hline 
$\vdots$&$\vdots$\\
\hline
$\msga_k^T$ & $\msgb_k^T$  \\
\hline 
$\msga^T P_1$ & $\msgb^T P_1$ \\
\hline 
$\msga^T P_2$ & $\msgb^T P_2$ \\
\hline 
$\vdots$&$\vdots$\\
\hline
$\msga^T P_r$ & $\msgb^T P_r$ \\
\hline 
\end{tabular}
\label{fig:regen1}
}\quad
\subfloat[Illustrating the piggybacking stated in Lemma~\ref{lem:regen1}. The parities $(k+1)$ and $(k+2)$ respectively correspond to $(k+\pa)$ and $(k+\pb)$ of the Lemma.]{
\begin{tabular}{|c|c|c|}
\hline 
$\msga_1^T$ & $\msgb_1^T$ \\
\hline 
$\vdots$&$\vdots$\\
\hline
$\msga_k^T$ & $\msgb_k^T$  \\
\hline 
$\msga^T P_1$ & $\msgb^T P_1 + \msga^T P_2 R$ \\
\hline 
$\msga^T P_2$ & $\msgb^T P_2$ \\
\hline 
$\vdots$&$\vdots$\\
\hline
$\msga^T P_r$ & $\msgb^T P_r$ \\
\hline 
\end{tabular}
\label{fig:regen2}
}\qquad
\subfloat[Piggybacking the `regenerating code' constructions of~\cite{cadambe2011permutation,ourITW, wang2011codes,papailiopoulos2011repair} for efficient parity repair]{
\begin{tabular}{c}
\normalnode 1  \\
$\vdots$\\
\normalnode k \\
\normalnode k+1\\
\normalnode k+2\\
\normalnode k+3 \\
\end{tabular}
\begin{tabular}{|c|c|c|}
\hline 
$\msga_1^T$ & $\msgb_1^T$ \\
\hline 
$\vdots$&$\vdots$\\
\hline
$\msga_k^T$ & $\msgb_k^T$  \\
\hline 
$\msga^T P_1$ \quad& $\msgb^T P_1+\msga^T P_2+\msga^T P_3$ \\
\hline 
$\msga^T P_2$ & $\msgb^T P_2$ \\
\hline 
$\msga^T P_3$ & $\msgb^T P_3$ \\
\hline 
\end{tabular}
\label{fig:regen3}
}
\label{fig:regen}
\caption{Piggybacking for efficient parity-repair in existing codes originally constructed for repair of only systematic nodes.}
\end{figure*}

Under a piggybacking as described in the lemma, the repair of parity node $(k+\pb)$ can be made more efficient by exploiting the fact that the parity node $(k+\pa)$ now stores the piggybacked symbol $(\msgb^T P_\pa+\msga^T P_\pb R)$.  We now demonstrate the use of this design by making the repair of parity nodes efficient in the explicit MDS `regenerating code' constructions of~\cite{cadambe2011permutation,ourITW, wang2011codes,papailiopoulos2011repair} which address the repair of only the systematic nodes. These codes have the property that \[ Q_\pa^{(i)} = Q_i \qquad \forall ~ i \in \{1,\ldots,k\},~~\forall~\pa \in \{1,\ldots,r\}\] i.e., the repair of any systematic node involves every parity node passing the same linear combination of its data (and this linear combination depends on the identity of the systematic node being repaired). It follows that in these codes, the condition~\eqref{eq:regen} is satisfied for every pair of parity nodes with $R$ and $S$ being identity matrices. 

\beg \label{eg:regen2}
The piggybacking of (two instances) of any such code~\cite{cadambe2011permutation,ourITW, wang2011codes,papailiopoulos2011repair}  is shown in Fig.~\ref{fig:regen3} (for the case $r=3$). As discussed previously, the MDS property and the property of efficient repair of systematic nodes is retained upon piggybacking. The repair of parity node $(k+1)$ in this example is carried out by downloading all the $2k\mu$ symbols. On the other hand, repair of node $(k+2)$ is accomplished by \accessing and downloading $\msgb$ from the systematic nodes, $(\msgb^T P_1 + \msga^T P_2 + \msga^T P_3)$ from the first parity node, and $(\msga^T P_3)$ from the third parity node. This gives the two desired symbols $\msga^T P_2$ and $\msgb^T P_2$. Repair of the third parity is performed in an identical manner, except that $\msga^T P_2$ is downloaded from the second parity node. The average amount of download and \access for the repair of parity nodes, as a fraction of the size $k\mu$ of the message, is thus
\[ \frac{2k+2}{3k} \]
which translates to a saving of around $33\%$.
\eeg

In general, the set of $r$ parity nodes is partitioned into
\[ g = \left\lfloor\frac{r}{\sqrt{k+1}} \right\rfloor \]
sets of equal sizes (or nearly equal sizes if $r$ is not a multiple of $g$). Within each set, the encoding procedure of Fig.~\ref{fig:regen3} is performed separately. The first parity in each group is repaired by downloading all the data from the systematic nodes. On the other hand, as in Example~\ref{eg:regen2}, the repair of any other parity node is performed by \accessing $\msgb$ from the systematic nodes, the second (which is piggybacked) symbol of the first parity node of the set, and the first symbols of all other parity nodes in the set. Assuming the $g$ sets have equal number of nodes (i.e., ignoring rounding effects), the average amount of \access and download for the repair of parity nodes, as a fraction of the size $k\mu$ of the message, is 
\[ \frac{1}{2} + \frac{k+(\frac{r}{g}-1)^2}{2k\left(\frac{r}{g}\right)}~.\]

\FloatBarrier
\section{Conclusions And Open Problems}\label{sec:conclusions}
We present a new \textit{piggybacking} framework for designing storage codes that require low data-\access and download during repair of failed nodes. This framework operates on multiple instances of existing codes and cleverly adds functions of the data from one instance onto  the other, in a manner that preserves properties such as minimum distance and the finite field of operation, while enhancing the repair-efficiency. We illustrate the power of this framework by using it to design the most efficient codes (to date) for three important settings.  
In the paper, we also show how this framework can enhance the efficiency of existing codes that focus on the repair of only systematic nodes, by piggybacking them to also enable efficient repair of parity nodes.

This simple-yet-powerful framework provides a rich design space for construction of storage codes. In this paper, we provide a few designs of piggybacking and specialize it to existing codes to obtain the four specific classes of code constructions. We believe that this framework has a greater potential, and clever designs of other piggybacking functions and application to other base codes could potentially lead to  efficient codes for various other settings as well. Further exploration of this rich design space is left as future work. Finally, while this paper presented only achievable schemes for data-\access efficiency during repair, determining the optimal repair-efficiency under these settings remains open. 

\bibliographystyle{IEEEtran}

\appendix
\begin{IEEEproof}[Proof of Theorem~\ref{thm:framework}]
Let us restrict our attention to only the nodes in set $S$, and let $|S|$ denote the size of this set. From the description of the piggybacking framework above, the data stored in instance $j~(1\leq j \leq \alpha)$ under the base code is a function of $U_j$. This data can be written as a $|S|$-length vector $\mathbf{f}(U_j)$ with the elements of this vector corresponding to the data stored in the $|S|$ nodes in set $S$. On the other hand, the data stored in instance $j$ of the piggybacked code is of the form $\left(\mathbf{f}(U_j)+\mathbf{g}_{j}(U_1,\ldots,U_{j-1})\right)$ for some arbitrary (vector-valued) functions `$\mathbf{g}$'. Now,
\bea I\left(\left\lbrace Y_i \right \rbrace_{i \in S};\ U_1,\ldots,U_{\alpha}\right)&=&
I\left(\left\lbrace \mathbf{f}(U_j)+\mathbf{g}_{j}(U_1,\ldots,U_{j-1}) \right \rbrace_{j=1}^\alpha\ ;\ U_1,\ldots,U_{\alpha}\right)\\
&=& \sum_{\ell=1}^{\alpha} I\left(\left.\left\lbrace \mathbf{f}(U_j)+\mathbf{g}_{j}(U_1,\ldots,U_{j-1}) \right \rbrace_{j=1}^\alpha;\ U_\ell \ \right|\ U_1,\ldots,U_{\ell-1}\right)\\
&=& \sum_{\ell=1}^{\alpha} I\left(\left.\mathbf{f}(U_\ell)\ ,\ \left\lbrace \mathbf{f}(U_j)+\mathbf{g}_{j}(U_1,\ldots,U_{j-1}) \right \rbrace_{j=\ell+1}^\alpha;\ U_\ell \ \right|\ U_1,\ldots,U_{\ell-1}\right)\\
&\geq& \sum_{\ell=1}^{\alpha} I\left(\left.\mathbf{f}(U_\ell)\ ;\ U_\ell \ \right|\ U_1,\ldots,U_{\ell-1}\right)\\
&=& \sum_{\ell=1}^{\alpha} I\left(\mathbf{f}(U_\ell)\ ;\ U_\ell\right)\\
&=&  I\left(\left\lbrace X_i \right \rbrace_{i \in S};U_1,\ldots,U_{\alpha}\right)~.
\eea
where the last two equations follow from the fact that the messages $U_\ell$ of different instances $\ell$ are independent.
\end{IEEEproof}

\end{document}